\documentclass[aps,prd,amsmath,floats,floatfix,
  superscriptaddress,
  nofootinbib,
  twocolumn,
  notitlepage,
  10pt,
  showpacs]{revtex4-1}
\usepackage{graphicx} 
\usepackage{subfigure}
\usepackage{amsmath,amssymb}
\usepackage{color}
\usepackage{xcolor}
\usepackage{hyperref}
\hypersetup{
    colorlinks,
    linkcolor={blue!20!black},
    citecolor={cyan!100!black},
    urlcolor={blue!20!black}
}
\usepackage[normalem]{ulem} 

\def\Lie{\mathcal{L}}
\def\p{\partial}

\begin{document}

\hfill {{\footnotesize USTC-ICTS/PCFT-24-54}}

\title{Boson star superradiance with spinning effects and in time domain}

\author{Fu-Ming Chang}
\email[]{changfum@mail.ustc.edu.cn}
\affiliation{Interdisciplinary Center for Theoretical Study, University of Science and Technology of China, Hefei, Anhui 230026, China}
\affiliation{Department of Modern Physics, University of Science and Technology of China, Hefei 230026, China}

\author{He-Yu Gao}
\email[]{gao\_he\_yu@mail.ustc.edu.cn}
\affiliation{Department of Modern Physics, University of Science and Technology of China, Hefei 230026, China}

\author{V\'ictor Jaramillo}
\email[]{jaramillo@ustc.edu.cn}
\affiliation{Department of Astronomy, University of Science and Technology of China, Hefei 230026, China}
\affiliation{Department of Modern Physics, University of Science and Technology of China, Hefei 230026, China}

\author{Xin Meng}
\email[]{mx15069132950@mail.ustc.edu.cn}
\affiliation{Department of Modern Physics, University of Science and Technology of China, Hefei 230026, China}

\author{Shuang-Yong Zhou}
\email[]{zhoushy@ustc.edu.cn}
\affiliation{Interdisciplinary Center for Theoretical Study, University of Science and Technology of China, Hefei, Anhui 230026, China}
\affiliation{Department of Modern Physics, University of Science and Technology of China, Hefei 230026, China}
\affiliation{Peng Huanwu Center for Fundamental Theory, Hefei, Anhui 230026, China}

\date{November 2024}
\begin{abstract}
Superradiance, the process by which waves are amplified through energy and angular momentum transfer, can also occur in horizonless objects like boson stars, due to both the real space and internal field space rotations. In this work, we study superradiance in the frequency and time domains for static and spinning boson stars, constructed within general relativity and with a self-interacting complex scalar field as a matter source. Using linear perturbation theory and three dimensional nonlinear simulations, we calculate amplification factors and analyze energy and angular momentum transfer in scattering processes, with results showing consistency between approaches. Wave scattering inside a cavity containing a boson star is also examined, demonstrating the effects of confinement on amplification.
\end{abstract}

\maketitle
\tableofcontents

\section{Introduction}

Under special circumstances, waves interacting with an object can be reflected in such a way that their amplitudes are amplified, transferring part of the object’s energy and/or angular momentum to the waves. This phenomenon, where radiation is enhanced, is known as superradiance. Superradiance was first studied by Ginzburg and Frank in the 1940s in the context of electromagnetic waves in continuous media \cite{Ginzburg:1945zz}. A form of superradiant amplification involving scalar and electromagnetic waves scattering off a rotating object was later proposed by Zeldovich \cite{Zeldovich:1971,Zeldovich:1972,Zeldovich:1986iw}. Since then, much of the research in this area has focused on rotational superradiance. This application is especially significant in relativity, where waves incident on a Kerr black hole can be superradiantly scattered if their frequency is sufficiently low \cite{Zeldovich:1971, Misner:1972kx,Press:1972zz,Starobinsky:1973aij,Starobinskil:1974nkd,Teukolsky:1974yv}, a phenomenon with notable astrophysical implications \cite{Cardoso:2004hs,Dolan:2007mj,Arvanitaki:2009fg,Bredberg:2009pv,Pani:2012vp,East:2013mfa,Richartz:2013unq,Berti:2015itd,East:2017ovw,Baryakhtar:2017ngi,Baumann:2018vus,Berti:2019wnn,Zhu:2020tht,Stott:2020gjj,Roy:2021uye,Chen:2022nbb,Siemonsen:2022yyf} (see \cite{Brito:2015oca} for a review on this topic).

Apart from rotational cases, superradiant phenomena can also arise in other scenarios. One well-known example occurs when waves interact with charged black holes \cite{Bekenstein:1973mi}. In the spherically symmetric Reissner-Nordström black hole, this theoretical setup allows for long-term, high-resolution evolution to track the nonlinear development of a superradiant instability \cite{Sanchis-Gual:2015lje,Bosch:2016vcp}. Another example is when superradiance emerges not from the physical rotation of the system, but from the internal rotation of scalar fields, as in the interaction of incident waves with horizonless objects like boson stars. This second scenario was first identified in the context of flat-spacetime Q-balls, where rotation in the complex field plane allows for energy and angular momentum transfer between coupled wave modes \cite{Saffin:2022tub}. It was shown that within a specific frequency range, the energy and angular momentum of the scattered waves can exceed those of the incident waves, leading to energy extraction and, consequently, superradiance. This initial analysis focused on static and spinning Q-balls in 2+1 dimensions and spherical Q-balls in 3+1 dimensions. Subtleties regarding the definitions of amplification factors when two modes with distinct propagating speeds are involved were clarified in \cite{Cardoso:2023dtm}. Subsequently, it was demonstrated that the more technically challenging cases of spinning Q-balls in 3+1 dimensions \cite{Zhang:2024ufh} and spherically symmetric boson stars with Q-ball-like scalar potentials \cite{Gao:2023gof} also exhibit superradiance. Moreover, energy amplification factors calculated using linear perturbation theory were shown to agree with those obtained from full nonlinear evolutions for the static non-spinning Q-ball in 2+1 dimensions \cite{Cardoso:2023dtm}. The existence of this form of superradiance in non-relativistic boson stars has also been briefly discussed in \cite{Cardoso:2023dtm}.

In this paper, we address the general phenomenon of superradiance when the scatterer is a spinning soliton coupled to gravity, specifically a rotating boson star. We analyze this scenario from both the frequency domain using linear perturbation theory and the time domain through nonlinear simulations. Stationary boson stars \cite{Kaup:1968zz,Ruffini:1969qy} are of particular interest in the context of scalar field dark matter models \cite{Matos:1998vk,Hu:2000ke,Suarez:2013iw, Hui:2016ltb}, as they are thought to form naturally through gravitational cooling processes \cite{Seidel:1993zk}. These objects also serve as practical examples of spacetimes where strong-gravity dynamical properties can be tested numerically \cite{Guzman:2005bs,Palenzuela:2006wp,Cunha:2017wao,Siemonsen:2024snb}. While the spherical boson star has been extensively studied, the literature also includes a variety of other configurations, such as rotating boson stars, along with investigations into their dynamics and stability in more specific cases (for reviews, see, \textit{e.g.}, \cite{Liebling:2012fv,Shnir:2022lba}).

Rotating boson stars \cite{Yoshida:1997qf}, unlike their spherical counterparts \cite{Gleiser:1988ih,Seidel:1990jh}, are found to be unstable in the minimal case of a massive scalar field without self-interactions. This instability manifests as a non-axisymmetric mode that grows over time \cite{Sanchis-Gual:2019ljs}. However, recent studies have shown that introducing nonlinear interactions in the bosonic field can stabilize these configurations \cite{Siemonsen:2020hcg,DiGiovanni:2020ror,Jaramillo:2024shi}. The existence of self-interacting rotating boson stars enables two significant advances. First, as revealed by our perturbative analysis, their nonlinear scalar potential allows for the extraction of energy from the star, even with perturbative waves. Second, these stabilized configurations make it possible to perform numerically robust simulations of wavepacket interactions.

The plan of this paper is as follows: In Sec.~\ref{sec:model}, we introduce the physical framework, consisting of a self-interacting complex scalar field minimally coupled to gravity. Sec.~\ref{sec:stationary} details the construction of spherical and spinning boson stars using a two-dimensional spectral solver, along with key diagnostic quantities and the fiducial configurations used in the analysis. In Sec.~\ref{sec:evol_scheme}, we describe the numerical relativity setup for evolving boson stars and interacting waves, covering both spherically symmetric and fully three-dimensional implementations. Sec.~\ref{sec:linear} presents the analysis of perturbative waves, including the computation of amplification factors for energy and angular momentum via a coupled system of partial differential equations. Sec.~\ref{sec:nonlinear} explores the full nonlinear scattering of wavepackets, starting with spherical configurations to validate the linear regime and examining the dynamics of scalar waves inside a cavity containing a boson star. We also investigate the non-spherical scenarios, presenting wave dynamics in three spatial dimensions and amplification factors in such spacetimes. Finally, Sec.~\ref{sec:conclusions} summarizes our findings, and the appendix provides validation tests.

\section{Model}\label{sec:model}

The field theory framework for this study is based on a complex scalar field $\Phi$ minimally coupled to gravity. The action governing the dynamics of the system is given by
\begin{equation}\label{eq:action}
S = \int \text{d}^4 \tilde{x}\sqrt{-g}\left[\frac{R}{16\pi}-\nabla^a\Phi^\dagger\nabla_a\Phi-V(|\Phi|)\right]\, ,
\end{equation}
where $g$ is the determinant of the spacetime metric $g_{ab}$, $R$ is the Ricci scalar, and $V(|\Phi|)$ is the scalar potential. The potential is chosen to respect the $\mathrm{U}(1)$ symmetry of the scalar field in the complex plane. For this work, we adopt the specific form
\begin{equation}\label{eq:potential}
V(|\Phi|) = \mu^2 |\Phi|^2 + \frac{\lambda}{2} |\Phi|^4 + \frac{\nu}{3} |\Phi|^6\, ,
\end{equation}
which includes mass $\mu$ and higher-order self-interaction terms.

Throughout this paper, we use natural units ($c = 1$, $G = 1$) and adopt the spacetime metric signature $(-,+,+,+)$. In these units, the scalar field $\Phi$ is dimensionless, while $\mu$, $\sqrt{\lambda}$, and $\sqrt{\nu}$ have dimensions of inverse length. To streamline the analysis and facilitate comparison with existing studies, we introduce dimensionless variables:
\begin{equation}
x^\mu = \tilde{x}^\mu\mu \, , \quad g = \frac{\lambda M_P^2}{\mu^2}\,, \quad h = \frac{\nu M_P^4}{\mu^2}\,,
\end{equation}
where $M_P = (8\pi)^{-1/2}$ is the reduced Planck mass. These rescaled quantities align with conventions used in earlier work, such as \cite{Gao:2023gof}. In some sections, we also redefine the scalar field as $f = \Phi/M_P$ for convenience, although we typically refer to $\Phi$ directly for clarity.

The equations of motion derived from the action are the Einstein-Klein-Gordon system:
\begin{equation}\label{eq:EoM}
{G^a}_b = 8\pi {T^a}_b, \quad \nabla_c\nabla^c\Phi - \frac{\text{d}V}{\text{d}|\Phi|^2}\Phi = 0,
\end{equation}
where ${G^a}_b$ is the Einstein tensor and ${T^a}_b$ is the energy-momentum tensor of the scalar field. The energy-momentum tensor is expressed as
\begin{equation}\label{eq:Tmunu}
T_{ab} = 2\nabla_{(a}\Phi^\dagger\nabla_{b)}\Phi - g_{ab}\left(\nabla^d\Phi^\dagger\nabla_d\Phi + V\right).
\end{equation}
The Einstein-Klein-Gordon system encapsulates (self-consistently) the coupling between spacetime geometry and the dynamics of the scalar field.

\section{The stationary configurations}\label{sec:stationary}

Spinning boson stars are stationary and axisymmetric solutions to the Einstein-Klein-Gordon system. We adopt the Lewis-Papapetrou coordinate system $\{t,r,\theta,\varphi\}$, in which the line element is given by:
\begin{equation}\label{eq:LP_metric}
\begin{split}
    ds^2 = &- e^{2 F_0} dt^2 + e^{2 F_1}\left(dr^2 + r^2 d\theta^2\right) \\
    &+ e^{2F_2}r^2\sin^2\theta\left(d\varphi - w dt\right)^2 \, ,
\end{split}
\end{equation}
where the functions $\{F_i\}$ and $w$ depending only on $r$ and $\theta$. The corresponding parametrization of the scalar field is
\begin{equation}\label{eq:ansatz_phi}
    \Phi_B = \phi(r,\theta) e^{-i(\omega_B t - m_B\varphi)} \, .
\end{equation}
where $\phi$ is a real function, $m_B$ (the harmonic azimuthal index) an integer and $\omega_B$ some positive number smaller than $\mu$. Spherical (static) boson stars are obtained by setting $m_B=0$, $w=0$, $F_1=F_2$ (for isotropic coordinates) while removing the $\theta$ dependence in the remaining functions. Substituting the ans\"atze from Eqs.~\eqref{eq:LP_metric} and \eqref{eq:ansatz_phi} into the system of equations \eqref{eq:EoM} leads to a system of partial differential equations involving five unknown fields. Following the approach of \cite{Gourgoulhon:2010ju, Grandclement:2014msa}, and introducing the shorthand notations  $N = e^{F_0}$, $A = e^{F_1}$, and $B = e^{F_2}$, the Einstein equations can be expressed in a remarkably compact form
\begin{subequations}\label{eq:Einstein}
\begin{align} 
\Delta_3 F_0 = &4\pi A^2(\rho + S)
	+ \frac{B^2 r^2\sin^2 \theta}{2N^2} \partial w\partial w \nonumber\\
	& - \partial F_0 \partial(F_0 + F_2) \, ,\\
\tilde \Delta_3 (w r \sin\theta) = &- 16\pi \frac{N A^2}{B^2} \frac{P_\varphi}{r\sin\theta}\nonumber\\
	& + r\sin\theta  \, \partial w \partial(F_0 - 3 F_2) \, ,\\
\Delta_2 \left[ (NB-1) r\sin\theta \right]
	&= 8\pi N A^2 B r\sin\theta (S^r_{\ \, r} + S^\theta_{\ \, \theta} ) \, ,\\
\Delta_2 (F_1 + F_0) &= 8\pi A^2 S^\varphi_{\ \, \varphi} 
 + \frac{3 B^2 r^2\sin^2 \theta}{4 N^2} \, \partial w\partial w\nonumber\\
	&- \partial F_0  \partial F_0  \, .
\end{align}
\end{subequations}
Where the following differential operators have been introduced:
\begin{subequations}\label{eq:operators}
\begin{eqnarray}
	&  & \Delta_2 := \partial_r^2 + \frac{1}{r}\partial_r
	+ \frac{1}{r^2}\partial_\theta^2 \, , \\
	& & \Delta_3 := \partial_r^2 + \frac{2}{r}\partial_r
	+ \frac{1}{r^2}\partial_\theta^2 + \frac{1}{r^2\tan\theta} \partial_\theta \, , \\
 	& & \tilde\Delta_3 := \Delta_3 - \frac{1}{r^2\sin^2\theta} \, , \\
    & & \partial f_1 \partial f_2 = \partial_r f_1 \partial_r f_2 + \frac{1}{r^2} \partial_\theta f_1 \partial_\theta f_2 \, .
\end{eqnarray}
\end{subequations}
The source terms, derived from Eq.~\eqref{eq:Tmunu} and using the 3+1 decomposition that will be discussed in the following section, are given by the expressions:
\begin{eqnarray}
    \rho + S &=& \frac{4}{N^2}\varrho^2\phi^2 - 2V \, ,\\
    P_\varphi &=& 2\frac{\varrho m_B\phi^2}{N} \, ,\\
    {S^r}_r + {S^\theta}_\theta &=& \frac{2}{N^2}\varrho^2\,\phi^2 - 2\frac{m_B^2}{B^2r^2\sin^2\theta}\phi^2 - 2V \, ,\\
    {S^\varphi}_\varphi &=& \frac{\varrho^2\,\phi^2}{N^2} + \frac{m_B^2}{B^2r^2\sin^2\theta}\,\phi^2 - \frac{\partial\phi\partial\phi}{A^2} - V ~~\,
\end{eqnarray}
with $\varrho = \omega_B-m_Bw$. The Klein-Gordon equation becomes also a compact expression for the scalar profile $\phi$ given by the following PDE:
\begin{equation}\label{eq:KleinGordon}
\begin{split}
    \Delta_3  \phi = &A^2\left(\frac{dV}{d|\Phi|^2}-\frac{\varrho^2}{N^2}\right)\phi \\
    &- \partial\phi\partial(F_0+F_2) + \frac{A^2}{B^2}\frac{m_B^2}{r^2\sin^2\theta}\phi \, .
\end{split}
\end{equation}

Spinning boson stars in the model \eqref{eq:action} are solitonic (particle-like) solutions and are required to be everywhere regular and asymptotically flat. This implies that the previous differential equations should be supplied with the following boundary conditions:
\begin{equation}\label{eq:out_bc}
  \begin{split}
    &\phi|_{r\to\infty}=0, \quad w|_{r\to\infty}=0 \, ;\\
    &F_0|_{r\to\infty}=0, \quad F_1|_{r\to\infty}=0, \quad F_2|_{r\to\infty}=0.
  \end{split}
\end{equation}
Regularity of the solution at the origin and on the symmetry axis require,
\begin{equation}\label{eq:regularity_r}
  \begin{split}
    &\phi|_{r=0}=0, \quad \partial_r w |_{r=0} = 0 \, ;\\
    &\partial_r F_0 |_{r=0}=0, \quad \partial_r F_1 |_{r=0}=0, \quad \partial_r F_2 |_{r=0}=0,
  \end{split}
\end{equation}
\begin{equation}\label{eq:regularity_th}
  \begin{split}
    &\phi|_{\theta=0,\pi}=0, \quad \partial_\theta w |_{\theta=0,\pi}=0 \, ;\\
    &\partial_\theta F_0 |_{\theta=0,\pi}=0, \quad \partial_\theta F_1 |_{\theta=0,\pi}=0, \quad \partial_\theta F_2 |_{\theta=0,\pi}=0,\\
    &F_1|_{\theta=0,\pi}=F_2 |_{\theta=0,\pi}.
  \end{split}
\end{equation}

The spacetime is expected to exhibit symmetry under reflection across the $\theta = \pi/2$ plane. For the scalar field profile, two types of solutions are possible: even and odd parity, as discussed in \cite{Kleihaus:2007vk}. However, in this work, we will focus on the even parity case. This implies that the derivatives of the metric functions and the scalar field $\phi$ with respect to $\theta$ vanish at $\theta = \pi/2$:
\begin{equation}\label{eq:reflection}
  \begin{split}
    & \partial_\theta\phi|_{\theta=\pi/2}=0, \quad \partial_\theta w |_{\theta=\pi/2}=0\, ;\\
    &\partial_\theta F_0 |_{\theta=\pi/2}=0, ~ \partial_\theta F_1 |_{\theta=\pi/2}=0, ~ \partial_\theta F_2 |_{\theta=\pi/2}=0 \, .
  \end{split}
\end{equation}

In summary, to construct a single boson star, we need to solve the elliptic system composed of Eqs.~\eqref{eq:Einstein} and \eqref{eq:KleinGordon}, while imposing the boundary conditions given by Eqs.~\eqref{eq:out_bc}-\eqref{eq:reflection}. These boundary conditions hold for every value of $m_B$, except for $m_B = 0$ (actually for $m_B = 1$, in addition to the boundary conditions discussed, special considerations must be applied to Eq.~\eqref{eq:KleinGordon}, as discussed in \cite{Grandclement:2014msa}). In the static, non-spinning spherical case ($m_B = 0$), we replace the boundary conditions for $\phi$ at the origin and along the symmetry axis with the following two requirements:
\begin{equation}
    \partial_r \phi|_{r=0}=0,\quad \partial_\theta \phi|_{\theta=\pi/2}=0 ~~ (m_B = 0) \,. 
\end{equation}
For both the spinning and the static cases, our approach is to fix the value of the lapse function $N$ at $r = 0$ and treat the frequency $\omega$ as a free parameter, which is varied alongside the other functions in an iterative Newton process. This process is carried out using the spectral solver provided by the \texttt{Kadath} library \cite{Kadath, Grandclement:2009ju}.

Stationary axisymmetric spacetimes possess two Killing vector fields which are, in coordinates \eqref{eq:LP_metric}
\begin{equation}
    \xi=\partial_t\, , ~~\chi=\partial_\varphi \, .
\end{equation}
The spacetime is asymptotically flat, allowing the total mass $M$ and angular momentum $L$ of the solutions to be computed using the Komar integrals, as outlined in \cite{Wald:1984rg}.
\begin{align}
    M&=\frac{1}{4\pi}\int_{\Sigma_t}{R_{ab}n^a\xi^b dV}\\
    \label{eq:globalquantM}
    &=2\int_{\Sigma_t}{\left(T_{ab}-\frac{1}{2}g_{ab}T\right)n^a\xi^b dV}, \\
    L&=-\frac{1}{8\pi}\int_{\Sigma_t}{R_{ab}n^a\chi^b dV}\\
    \label{eq:globalquantJ}
    &=-\int_{\Sigma_t}{T_{ab}n^a\chi^b dV} \, .
\end{align}
Here, $\Sigma_t$ represents a spacelike surface, and $n^\mu$ is a unit vector normal to $\Sigma_t$. In both equations, the Einstein equations have been applied, and in the final equality, we used the fact that $\xi$ and $\chi$ are orthogonal. These mathematical objects will be described in greater detail in Sec.~\ref{sec:evol_scheme}. For spherical boson stars, we also employ a standard definition of the radius to characterize their size. This is given by the radius $R_{99}$, which denotes the areal radius that encloses 99\% of the total boson star mass. Specifically, it is defined as the radius where the Misner-Sharp function \cite{Misner:1964je} equals 99\% of the total mass $M$.

In addition, the action \eqref{eq:action} possesses a global $U(1)$ symmetry. The current associated to this symmetry is
\begin{equation}\label{eq:N_current}
    j_a=i\left(\Phi^\dagger\partial_a\Phi-\Phi\partial_a\Phi^\dagger\right)\, ,
\end{equation}
which is divergence-free. Integration of it into $n^c$ over $\Sigma_t$ gives the conserved Noether charge
\begin{equation}
\label{eq:Q}
    Q=\int_{\Sigma_t} {j^a n_a dV} \, .
\end{equation}
It can be readily shown that, under the assumptions made for the scalar field and the spacetime metric, the ``quantization'' law for the angular momentum of boson stars, $L = m_B Q$, holds, as demonstrated in \cite{Schunck1996}.

\begin{figure}
\centering
\subfigure[~~Static]{
    \includegraphics[width=0.48\textwidth]{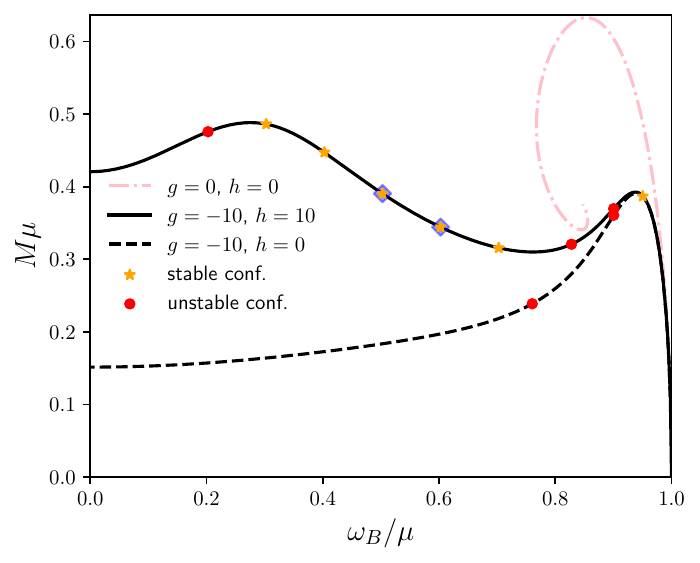}
}
\subfigure[~~Spinning ($m_B=1$)]{
    \includegraphics[width=0.48\textwidth]{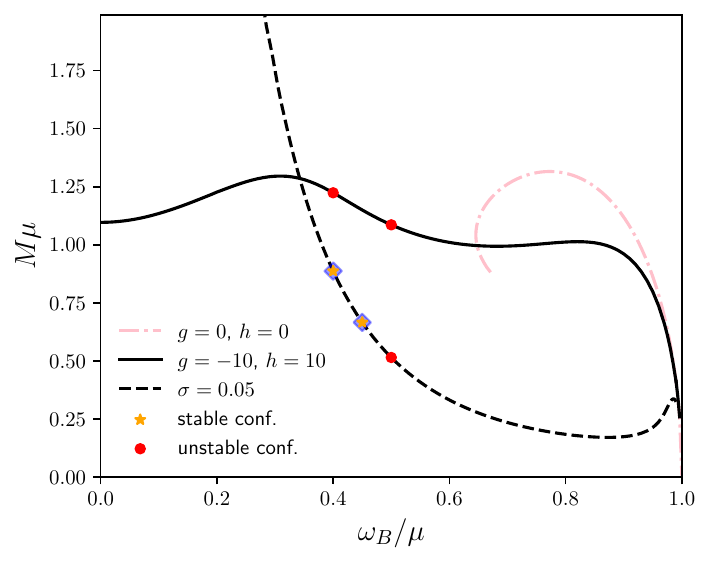}
}
  \caption{
  Sequences of spherical and spinning boson stars with different values of the self-interaction parameters $g$ and $h$.
  The case labeled with $\sigma = 0.05$ in the panel (b) corresponds to a sextic potential \eqref{eq:solitonicV} with values $g=-127.3$ and $h=3039.6$.
  The four stable configurations marked with blue diamonds are the ones used for calculating amplification factors and doing evolutions throughout the next sections of the paper.
  }
  \label{fig:M-spherical}
\end{figure}

Now we analyze explicit solutions of boson stars with a scalar potential containing the quartic and sextic terms. In Fig.~\ref{fig:M-spherical} panel (a), we present three families of spherical boson star solutions, each with different values of $g$ and $h$. While a more detailed analysis of the quartic and sextic terms could be conducted, the displayed families provide a clear illustration of the effects of including these interactions in the model.

Initially, we observe that the mini boson star family (dashed pink curve) aligns with the other two families near the Newtonian limit $(\omega_B \to \mu, M \to 0)$. This is because these configurations are more dilute, reducing the influence of self-interacting terms. As the scalar field amplitude increases, particularly near the center of the star, the quartic term $g$ becomes significant. Its negative value introduces an attractive self-interaction, which enhances gravitational attraction and limits the amount of mass the scalar field can support. As a result, the solutions become less massive. This behavior aligns with the findings in \cite{Barranco:2010ib}, which reported similar effects for a self-gravitating system composed of axions.

As the scalar field continues to increase in magnitude and we move toward smaller values of $\omega$, the positive sextic term $h$ becomes significant, introducing a repulsive interaction that allows for higher mass values. This is an important characteristic, as it leads to the existence of a \textit{second branch} of solutions where $dM/d\omega < 0$ before reaching the minimum $\omega$ turning point. For the spherical case, this second branch is found to be stable. Additionally, we will find that in this region, the amplification factors are notably enhanced compared to the first stable branch, which connects to the Newtonian limit.

We will return to Fig.~\ref{fig:M-spherical} to further discuss the stability properties and examine the selected solutions for analysis in the following sections. The construction of spherical boson stars is performed using both the shooting algorithm described in \cite{Gao:2023gof} and the spectral code presented in \cite{Alcubierre:2021psa}. In the spectral code, we impose $F_2 = F_1$ and $w = 0$ in Eq.~\eqref{eq:LP_metric}, while in the shooting code, the problem is solved using the polar-areal gauge.

Panel (b) of Fig.~\ref{fig:M-spherical} illustrates examples of spinning boson star sequences with non-zero self-interactions. These solutions differ qualitatively from the static case, not only due to rotation but also because of their toroidal energy (and charge) density morphology \cite{Yoshida:1997qf}. Despite this, the effect of the self-interacting potential on classical global quantities remains similar to that observed in the spherical case.
\begin{figure*}
\centering
\subfigure[~~Real part and imaginary part of $\Phi$ on the $xy$ plane]{
    \includegraphics[width=0.3\textwidth]{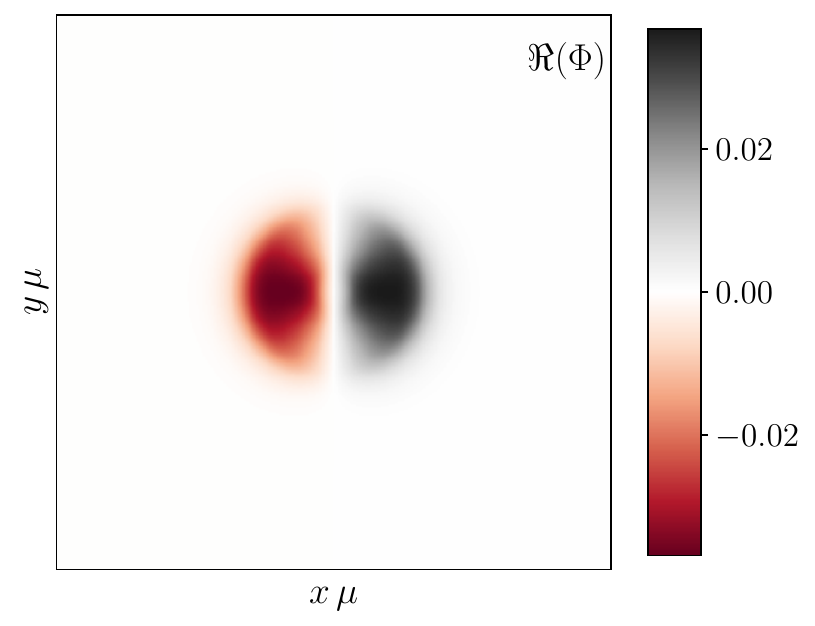}
    \includegraphics[width=0.3\textwidth]{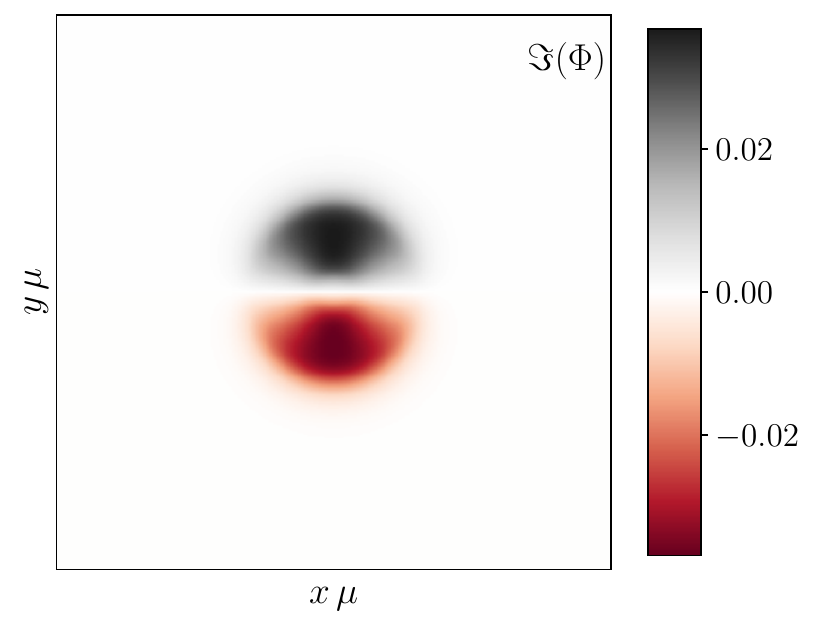}}
\subfigure[~~Magnitude of $\Phi$ on the $xy$ plane]{
    \includegraphics[width=0.3\textwidth]{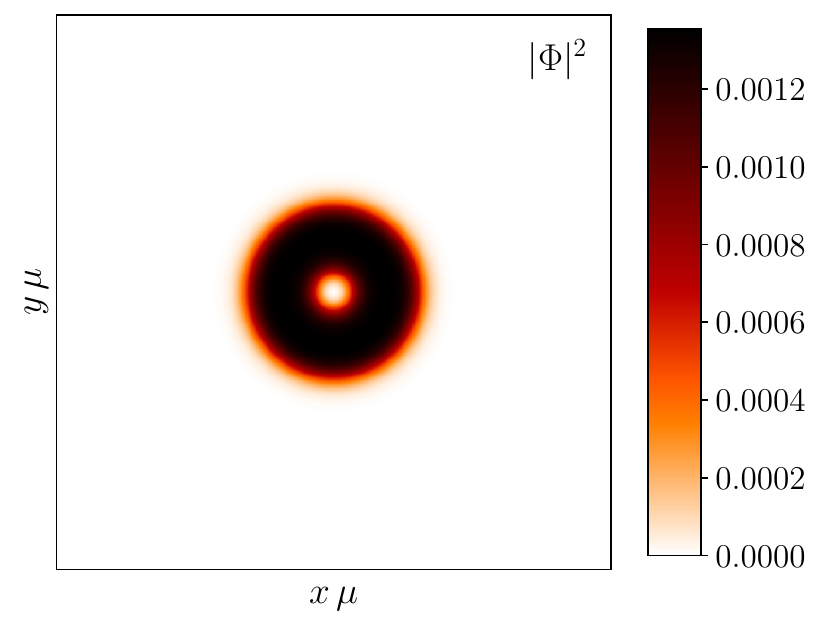}}\\
\subfigure[~~Metric function $e^{F_0}$ on the $xy$ and $xz$ planes]{
    \includegraphics[width=0.3\textwidth]{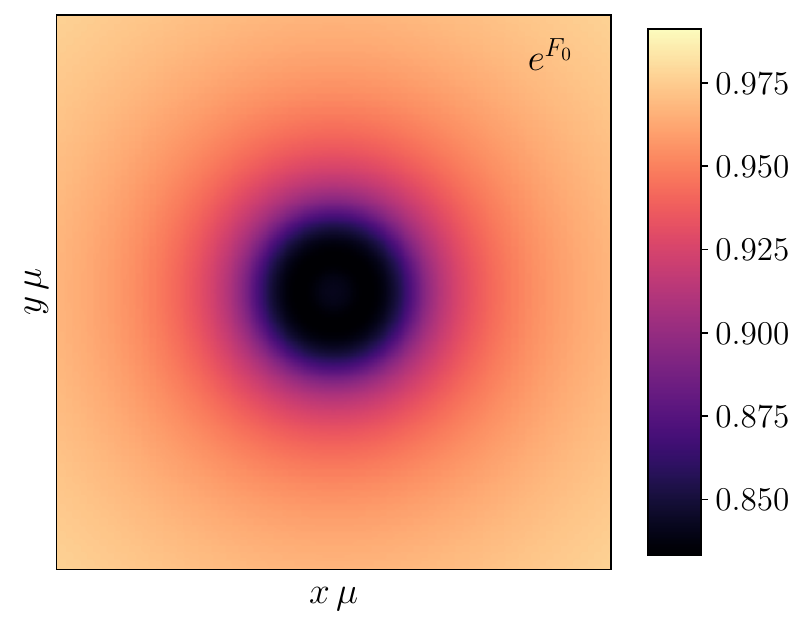}
    \includegraphics[width=0.3\textwidth]{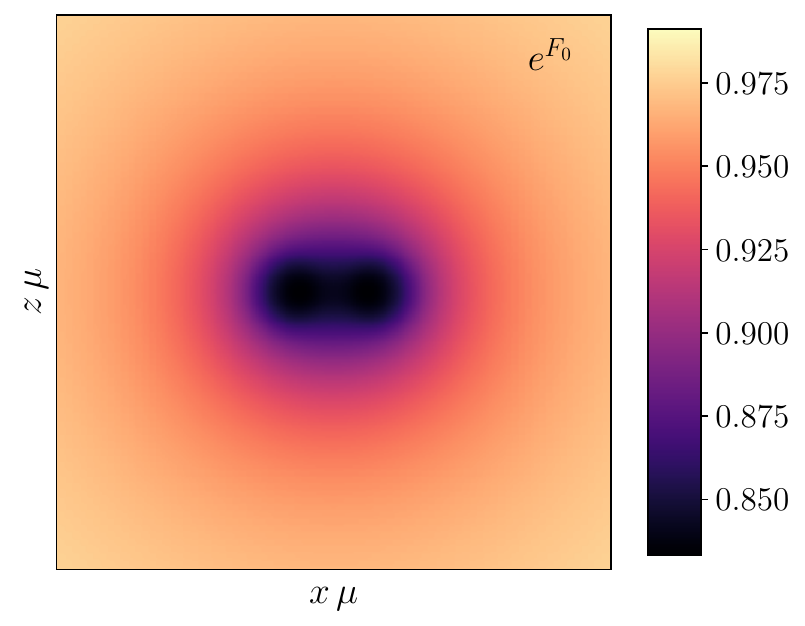}}
\subfigure[~~Magnitude of $\Phi$ on the $xz$ plane]{
    \includegraphics[width=0.3\textwidth]{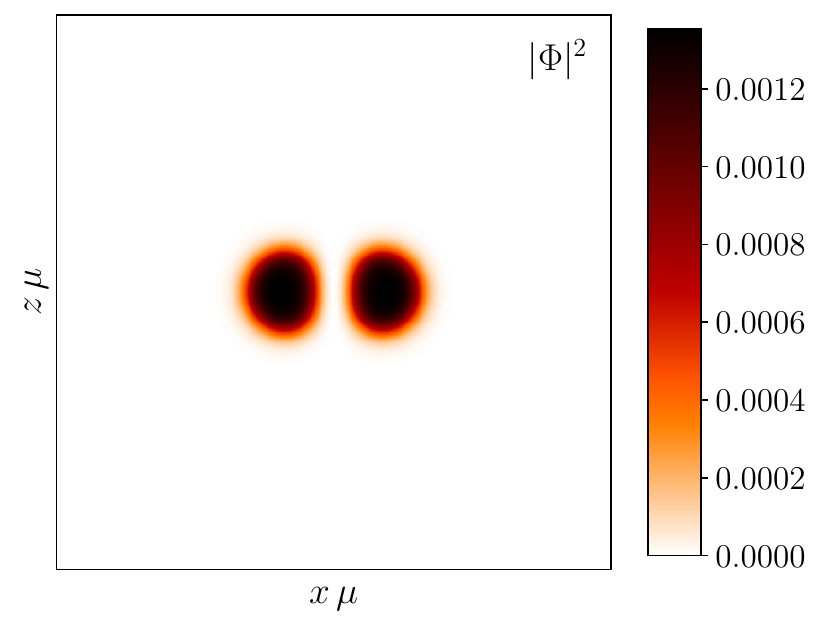}
}
  \caption{
    Illustrative solitonic potential spinning boson star with $\omega_B = 0.45\mu$ and $\sigma=0.05$. Panel (a): The real and imaginary parts are plotted at $t=0$. They rotate anti-clockwise, according to Eq.~\eqref{eq:ansatz_phi}. Panel (b) and (d): The magnitude of the scalar field on the $xy$ and $xz$ planes. Panel (c): Lapse function $e^{F_0}$ in both planes. The side of the box in all cases is $20/\mu$.
  }
  \label{fig:spinning_contours}
\end{figure*}

\begin{figure*}
\centering
    \includegraphics[width=0.6\textwidth]{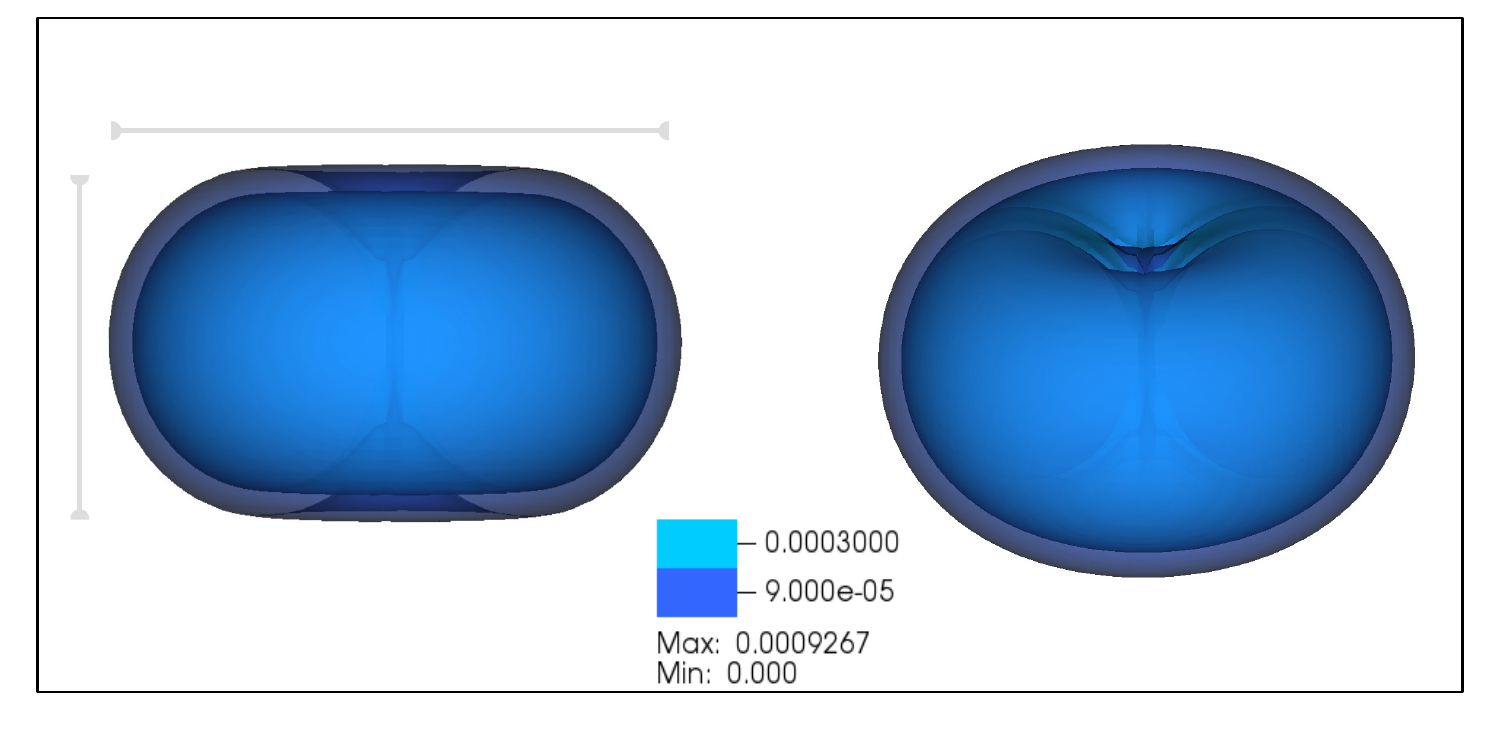}
  \caption{
  A couple of surfaces of constant energy density $\rho$ from two different perspectives. The horizontal and vertical lines in the left panel have lengths of $15/\mu$ and $9/\mu$ respectively. The configuration corresponds to the same star as in Fig.~\ref{fig:spinning_contours}.
  }
  \label{fig:toro}
\end{figure*}

This conclusion applies to stationary, isolated, unperturbed configurations, but the situation changes significantly when these solutions are studied dynamically. The branches identified as stable in the static $m_B = 0$ case generally correspond to unstable configurations when rotation is introduced. In fact, the entire sequence of spinning solutions becomes unstable against nonaxisymmetric perturbations, as noted in \cite{Sanchis-Gual:2019ljs}. Motivated by recent findings \cite{DiGiovanni:2020ror, Siemonsen:2020hcg}, we consider specific values of $g$ and $h$ derived from the ``solitonic potential'':
\begin{equation}\label{eq:solitonicV}
V = \mu^2 |\Phi|^2 \left(1 - 2\frac{|\Phi|^2}{\sigma^2}\right)^2 \, ,
\end{equation}
where $\sigma$ is a constant. For the spinning $m_B = 1$ stars, we will explore two cases: one using the polynomial potential \eqref{eq:potential} with $g = -10 = -h$, and another using the solitonic potential with $\sigma = 0.05$. The details of these models will be revisited in the following sections, however to exemplify the kind of solution a spinning boson star is, in Fig.~\ref{fig:spinning_contours} we have plotted some relevant functions of an illustrative solution with $\sigma = 0.05$, $\omega_B = 0.45\mu$. And, for the same configuration we show the toroidal topology of the surfaces of constant energy density. For $m_B=1$ stars, the energy density $\rho$ is nonzero at the symmetry axis \cite{Collodel:2019ohy}, as shown in Fig.~\ref{fig:toro}, although the scalar field vanishes [see e.g., Fig.~\ref{fig:spinning_contours} panels (b) and (d)].

The final point we wish to address concerning the stationary configurations under consideration involves the energy conditions for the fiducial cases. The model described by Eq.~\eqref{eq:action} consistently satisfies the null energy condition, regardless of the parameter choices. This condition requires that for any null vector $k^a$, the inequality
\begin{equation}
    T_{ab}k^a k^b \geq 0
\end{equation} 
must be satisfied. Our analysis confirms that this holds true for all choices of the potential $V$.

On the other hand, the weak energy condition, which requires that for any timelike vector $v^a$, the inequality 
\begin{equation}
    T_{ab}v^a v^b \geq 0
\end{equation} 
holds, is not always satisfied in these models. This is due to the potential in Eq.~\eqref{eq:potential} not being positive definite (since $g < 0$). The rotating fiducial family we consider has a positive definite potential, ensuring that the weak energy condition is met. However, this is not true for the spherical configurations. In the spherical case, the energy-momentum tensor is of the first Segrè type, meaning the weak energy condition is satisfied only if the null energy condition is met and the energy density measured by a static observer is positive everywhere \cite{Hawking:1973uf,Westmoreland:2013zxw}.

We have confirmed that this condition holds for the two fiducial $m_B = 0$ cases. However, within this same family of solutions (with $g = -10, h = 10$), all configurations in the range $\omega_B/\mu \in [0, 0.398]$ violate the weak energy condition. Notably, the solution with $\omega_B = 0$, which corresponds to a real scalar field, avoids Derrick’s argument (in curved spacetime \cite{Diez-Tejedor:2013sza}) because it does not satisfy the weak energy condition.

\section{Evolution scheme(s)}
\label{sec:evol_scheme}

We use the 3+1 decomposition of the spacetime metric,
\begin{equation}\label{eq:3+1_metric}
ds^2 = -\left(\alpha^2 - \beta_i\beta^i\right) dt^2 + 2\beta_i dtdx^i + \gamma_{ij} dx^i dx^j \, ,
\end{equation}
to reformulate the Einstein-Klein-Gordon system as a Cauchy problem. This approach foliates the spacetime into spacelike hypersurfaces $\Sigma_t$ of constant $t$, where points within each $\Sigma_t$ are labeled by spatial coordinates $x^i$. Using the Einstein equations, the evolution system for the spatial metric $\gamma_{ij}$ and the extrinsic curvature $K_{ij} = - \frac{1}{2\alpha} \left( \partial_{t} - \mathcal{L}_{\beta} \right) \gamma_{ij}$, where $\mathcal{L}_{\beta}$ is the Lie derivative along the shift vector $\beta$, can then be rewritten in the following form
\begin{subequations}\label{eq:dt}
\begin{align}
  \p_{t} \gamma_{ij} & = - 2 \alpha K_{ij} + \Lie_{\beta} \gamma_{ij} \,,
                       \label{eq:dtgamma} \\
  \p_{t} K_{ij}      & =  - D_{i} \p_{j} \alpha
                       + \alpha \left( R_{ij} - 2 K_{ik} K^{k}{}_{j} + K K_{ij} \right) \nonumber \\
                       & \quad + \Lie_{\beta} K_{ij} 
                       + 4\pi \alpha \left[ (S-\rho) \gamma_{ij} - 2 S_{ij} \right] \,,
                                         \label{eq:dtKij}
\end{align}
\end{subequations}
%
where $D_i$ is the covariant derivative with respect to the $3$-metric and the gravitational sources are given by the following projections of the energy-momentum tensor $T_{\mu\nu}$:
\begin{equation}
  \label{eq:source}
   \begin{aligned}
  \rho & \equiv T^{\mu \nu}n_{\mu}n_{\nu} \,,\\
  P_i  &\equiv -\gamma_{i\mu} T^{\mu \nu}n_{\nu} \,, \\
  S_{ij} &\equiv \gamma^{\mu}{}_i \gamma^{\nu}{}_j T_{\mu \nu} \,, \\
  S     & \equiv \gamma^{ij}S_{ij} \,.
   \end{aligned}
\end{equation}
Here, $n^\mu$ is the normal vector to the surfaces $\Sigma_t$ such that $n_\mu n^\mu=-1$. 

In addition to Eqs.~\eqref{eq:dtgamma} and \eqref{eq:dtKij}, the following set of constraints must be satisfied at all times
\begin{align}
\label{eq:Hamiltonian}
\mathcal{H} & \equiv {}^{(3)}R - K_{ij} K^{ij} + K^2 - 16 \pi \rho
       = 0\,,\\
\label{eq:momentumConstraint}
\mathcal{M}_{i} & \equiv D^{j} K_{ij} - D_{i} K 
        - 8\pi P_i
       = 0 \,.
\end{align}

Although simulations involving spherical stars represent a specific case of the rotating scenario, where $m_B = 0$, we have treated the evolutions of both cases separately. This allows spherical evolutions to be performed with greater precision and in less computational time.

It is important to note that the above equations are expressed as true tensors, meaning they can be applied using spherical coordinates, reduced to spherical symmetry, or used in the full 3D case with Cartesian coordinates. In the following sections, we will outline the details of the spherical code used, including its regularization procedure, as well as the 3D code and the hyperbolic formulation employed for that case.

\subsection{Spherical simulations} \label{sec:evol_scheme_sph}
For the spherical case, we used a strongly hyperbolic formulation that is regular at the origin\footnote{In parallel, we compared our results with those obtained using a code that employs the BSSNOK formulation, obtaining consistent results (see Appendix \ref{app:tests}).}.

To this end, we have employed the formulation presented for an application in spherical symmetry in the numerical relativity book \cite{alcubierre2008introduction} and used, for example, in \cite{Berczi:2021hdh}. Choosing spherical coordinates in Eq.~\eqref{eq:3+1_metric} $x^i = (r,\theta,\varphi)$, assuming spherical symmetry and choosing $\beta^i=0$, we can write the metric \eqref{eq:3+1_metric} as 
\begin{equation}
d s^{2}=-\alpha^{2}(t,r)d t^{2}+a(t,r)d r^{2}+r^{2}b(t,r)(d\theta^{2}+\sin^{2}\theta d\varphi^{2}) \, .
\end{equation}

To cast the Einstein-Klein-Gordon equations as first order equations, we can define the following variables
\begin{subequations}
\begin{eqnarray}
&{{K_{a}=-\frac{1}{2\alpha}\frac{\dot{a}}{a},~~~~~~D_{a}=\frac{a^{\prime}}{a},}}\\ &{{K_{b}=-\frac{1}{2\alpha}\frac{\dot{b}}{b},~~~~~~D_{b}=\frac{b^{\prime}}{b},}}\\&{\lambda={\frac{1}{r}}\left(1-{\frac{a}{b}}\right),\quad D_{\alpha}={\frac{\alpha^{\prime}}{\alpha}}}\\
&\Psi=\Phi^{\prime},~~~~~~\Pi=\frac{a^\frac{1}{2}b}{\alpha}\dot{\Phi}
    \end{eqnarray}
\end{subequations}
where the dot represents differentiation with respect to $t$, and the prime represents differentiation with respect to $r$.

Furthermore, we must perform some other variable changes to ensure the evolution equations are strongly hyperbolic. Namely, we will use $(K,{\bar{U}})$ instead of $(K_{a},D_{a})$
\begin{subequations}
\begin{eqnarray}
    &{{K=K_{a}+2K_{b},}}\\
    &{{\tilde{U}=D_{a}-2D_{b}-\frac{4b\lambda}{a}.}}
    \end{eqnarray}
\end{subequations}

In spherical symmetric case, the constraints \eqref{eq:Hamiltonian} and \eqref{eq:momentumConstraint} become
\begin{align}
\label{eq:spherical_momentumConstraint}
        &{{\mathcal{M}_r=2K_{b}^{\prime}-\left(\frac{2}{r}+{D_{b}}\right)(K-3K_{b})+{8\pi P_r}=0}}\\ 
\label{eq:spherical_Hamiltonian}
    \begin{split}
        &\mathcal{H}=2D_{b}^{\prime}+\frac{2}{r}\left(\lambda+D_{b}-\tilde{U}-\frac{4\lambda b}{a}\right)-D_{b}\left(\frac{1}{2}D_{b}\right.\\
        &\qquad\left.+\tilde{U}+\frac{4\lambda b}{a}\right) -aK_b(4K-6K_b)+16\pi a\rho=0
        \end{split}
\end{align}

And evolution equations are listed as follows.
\begin{subequations}\label{eq:evol_spherical}
    \begin{flalign}
        & \dot{\Phi}=\frac{\alpha}{a^{\frac{1}{2}} b} \Pi, \\
        & \dot{\Psi}=\partial_r\left(\frac{\alpha}{a^{\frac{1}{2}} b} \Pi\right), \\
        & \dot{\Pi}=\frac{1}{r^2} \partial_r\left(\frac{\alpha b r^2}{a^{\frac{1}{2}}} \Psi\right)-\alpha b a^{\frac{1}{2}} \frac{dV}{d|\Phi|^2}\Phi,\\
                &\dot{a}  =-2 \alpha a\left(K-2 K_b\right), \\
                &\dot{b}  =-2 \alpha b K_b, \\
                &\dot{D}_b  =-2 \partial_r\left(\alpha K_b\right) ,\\
                    &\dot{\alpha}  =-\alpha^2 f(\alpha) K \\
                    &\dot{D}_\alpha  =-\partial_r(\alpha f(\alpha) K),\\
&\dot{K}_b=  \frac{\alpha}{a r}\left[\frac{1}{2} \tilde{U}+\frac{2 \lambda b}{a}-D_b-\lambda-D_\alpha\right]\nonumber \\
                        & \qquad+\frac{\alpha}{a}\biggl[-\frac{1}{2} D_\alpha D_b-\frac{1}{2} D_b^{\prime}+\frac{1}{4} D_b\left(\tilde{U}+\frac{4 \lambda b}{a}\right)\nonumber\\
                        & \qquad+a K K_b\biggr]+4\pi{\alpha}\left(S^r\,_r-\rho\right),\\
                            &\dot{K}=  \alpha\left(K^2-4 K K_b+6 K_b^2\right) \nonumber\\
                            &\qquad-\frac{\alpha}{a}\left[D_\alpha^{\prime}+D_\alpha^2+\frac{2 D_\alpha}{r}-\frac{1}{2} D_\alpha\left(\tilde{U}+\frac{4 \lambda b}{a}\right)\right]\nonumber \\
                            &\qquad +4\pi{\alpha}\left(\rho+S^r\,_r+2 S^\theta\,_\theta\right) ,\\
&\dot{\lambda}=\frac{2 \alpha a}{b}\left[K_b^{\prime}-\frac{1}{2} D_b\left(K-3 K_b\right)+4\pi{P_r}\right],\\
    &\dot{\tilde{U}}=  -2 \alpha\biggl[K^{\prime}-2\left(K-3 K_b\right)\left(D_b-\frac{2 \lambda b}{a}\right)\nonumber\\
    &\qquad+D_\alpha\left(K-4 K_b\right)\biggr] -32\pi \alpha {P_r},
    \end{flalign}
\end{subequations}
where $f(\alpha)$ is an arbitrary function related to gauge in slicing. To avoid singularities, we choose $f(\alpha)=2/\alpha$, \textit{i.e.}, the 1+log gauge condition~\cite{Bona:1994dr}.

We use formulas \eqref{eq:evol_spherical} to simulate the evolution. In our code, we use 4th order implicit Runge–Kutta method. To compute the spacial derivative, we use 4th order finite differential method. To prevent the divergence, we add 5th order Kreiss-Oliger dissipation as described in \cite{Berczi:2021hdh}. For each time step$\phi^{n+1}=\phi^{n}+dtG(\phi^{n})$, the dissipation term is $G(\phi^{n}) \rightarrow G^{\prime}(\phi^{n}) = G(\phi^{n}) - \epsilon(-1)^{N}dr^{2N-1}\partial_{r}^{2N}\phi^{n}$, here $N$ is integer $N\geq1$. 

\subsection{3D simulations}\label{sec:evol_scheme_3D}

For evolving the axisymmetric scenarios, we use the Baumgarte-Shapiro-Shibata-Nakamura (BSSN) formulation~\cite{Nakamura:1987zz,Shibata:1995we,Baumgarte:1998te}. In this approach, similar to the extrinsic curvature for the spacetime metric, we make use of the ``canonical momentum'' of the complex scalar field $\Phi$, defined as:
\begin{equation}
\label{eq:Kphi}
K_{\Phi} = -\frac{1}{2\alpha}  \left( \partial_{t} - \Lie_{\beta} \right) \Phi \,,
\end{equation}
in which case the following first order evolution equations for the scalar field are obtained,
\begin{subequations}\label{eq:dt2}
    \begin{align}
  \p_{t} \Phi & = - 2 \alpha K_\Phi + \Lie_{\beta} \Phi\
                \,, \label{eq:dtPhi} \\
  \p_{t} K_\Phi &  = \alpha \biggl[ K K_{\Phi}
                  + \frac{1}{2} \left(\mu^2+\lambda|\Phi|^2+\nu|\Phi|^4\right) \Phi \nonumber \\
                 & \quad - \frac{1}{2} \gamma^{ij} D_i \partial_j \Phi\biggr] - \frac{1}{2} \gamma^{ij} \partial_i \alpha \partial_j \Phi
                       + \Lie_{\beta} K_\Phi \,, \label{eq:dtKphi}
    \end{align}
\end{subequations}
The real and imaginary part of the scalar field $\Phi$ 
are evolved separately inside the code.

To rewrite the Einstein equations in the BSSNOK formulation (see \cite{alcubierre2008introduction} for details), several auxiliary variables are introduced. One key element is the conformal metric with a unit determinant, defined as
\begin{equation}
{\tilde{\gamma}}_{ij} = \chi\,\gamma_{ij}, \quad {\tilde{\gamma}}^{ij} = \frac{1}{\chi}\,\gamma^{ij},
\end{equation}
where $\chi$ is a conformal factor.

By reformulating the Einstein equations using these variables, along with additional auxiliary fields, we obtain a system of equations governing both the evolution of the gravitational sector and the scalar field. Ultimately, the reformulated version of the evolution equations \eqref{eq:dt} and \eqref{eq:dt2}, along with the equations of motion for the auxiliary variables, lead us to the complete set of dynamical equations for the gravitational field and the complex scalar field in the BSSNOK framework:
\begin{subequations}
\label{eq:BSSNfull}
\begin{eqnarray}
\left( \partial_t -  \mathcal{L}_\beta \right)& \tilde \gamma_{ij} & = 
        - 2 \alpha \tilde A_{ij}\, , \\
\left( \partial_t -  \mathcal{L}_\beta \right)& \chi  & = 
        \frac{2}{3} \alpha \chi K\, , \\
\left( \partial_t -  \mathcal{L}_\beta \right)& K & = 
        [\dots] + 4 \pi \alpha (\rho + S)\, , \\
\left( \partial_t -  \mathcal{L}_\beta \right)& \tilde A_{ij} & = 
        [\dots] - 8 \pi \alpha \left(
          \chi S_{ij} - \frac{S}{3} \tilde \gamma_{ij}
        \right)\, , \\
\left( \partial_t -  \mathcal{L}_\beta \right)& \tilde \Gamma^i & = 
        [\dots] - 16 \pi \alpha \chi^{-1} P^i\, ,\\
  \left(\p_t - \Lie_{\beta} \right)& \Phi & = - 2 \alpha K_\Phi \,, \\
  \left(\p_t - \Lie_{\beta} \right)& K_\Phi &  = \alpha \left[ K K_{\Phi} - \frac{1}{2} \chi\tilde{\gamma}^{ij} \tilde{D}_i \partial_j \Phi \right.  \nonumber \\
  & & + \frac{1}{2}\left( \mu^2 + \lambda|\Phi|^2 + \nu|\Phi|^4 \right)\Phi \nonumber\\
  & & \left. + \frac{1}{4} \tilde{\gamma}^{ij} \partial_i \Phi \partial_j\chi\right] - \frac{\chi}{2} \tilde{\gamma}^{ij} \partial_i \alpha \partial_j \Phi \,. ~~
\end{eqnarray}
\end{subequations}

In the evolution equations for K (the trace of the extrinsic curvature), $\tilde{A}_{ij}$ (the traceless part of the conformal extrinsic curvature), and $\tilde{\Gamma}^i$ (the conformal connection functions), the terms denoted by $[\dots]$ represent the standard right-hand side of the BSSNOK equations in the absence of matter source terms. Here, $\tilde{D}$ refers to the covariant derivative compatible with the conformal metric $\tilde{\gamma}$.

The system of equations \eqref{eq:BSSNfull} is solved using the following procedure: The initial data is first set up and then evolved within the open-source \texttt{Einstein Toolkit} infrastructure \cite{EinsteinToolkit:2023_11,Loffler:2011ay}. The metric quantities are evolved using the \texttt{McLachlan} thorn \cite{Brown:2008sb} with the gauge conditions 1+log for $\alpha$ \cite{Bona:1994dr} and the ``Gamma-driver'' condition for $\beta^i$ \cite{Alcubierre:2002kk}. A modified version of the \texttt{Scalar} thorn, as used in \cite{Jaramillo:2024smx} and based on \cite{Cunha:2017wao,Canuda:zenodo}, is employed for evolving the scalar field. To monitor the formation of apparent horizons during the simulation, the \texttt{AHFinderDirect} thorn \cite{Thornburg:2003sf} is utilized. The evolution process is performed with fixed refinement levels, managed by the \texttt{Carpet} thorn \cite{Schnetter:2003rb}, which supports efficient grid refinement.

\section{Perturbative waves and amplification factors}\label{sec:linear}

We now address the calculation of energy and angular momentum amplification or attenuation during the scattering of waves off a boson star with arbitrary $m_B$. This analysis is performed in the linear regime, serving as a reference for the nonlinear evolution of wavepackets, which will be discussed in the next section. The case of spherically symmetric perturbative waves scattering off a $m_B = 0$ boson star was previously studied in \cite{Gao:2023gof}, and we use this scenario for consistency checks. The focus here, however, is on the scattering of nonspherical waves from spherical boson stars and, more generally, spinning boson stars. We will present results for these cases in the following.

We begin by considering a small perturbation $\Theta$ on top of a spinning boson star background~\eqref{eq:ansatz_phi},
\begin{equation}
\Phi\left(  t,r,\theta\right)  =\Phi_{B}\left(  t,r,\theta\right)
+\Theta\left(  t,r,\theta\right) \, . \label{perturbation}
\end{equation}
The complete field $\Phi$ is controlled by the equation of motion \eqref{eq:EoM}, so expanding the term $dV/d|\Phi|^2\Phi = \partial V/\partial\Phi^{\dag}$ near $\Phi_{B}$, we have
\begin{equation}
\frac{\partial V}{\partial\Phi^{\dag}}=\left.\frac{\partial V}{\partial\Phi^{\dag}
}\right|_{\Phi_{B}}+\left.\frac{\partial^{2}V}{\partial\Phi^{\dag}\partial\Phi}\right|_{\Phi
_{B}}\Theta+\left.\frac{\partial^{2}V}{\partial\Phi^{\dag}\partial\Phi^{\dag}
}\right|_{\Phi_{B}}\Theta^{\dag} \, .
\end{equation}
To leading order, the equation of motion for the perturbation is 
\begin{equation}\label{E.O.M}
\nabla_{\mu}\nabla^{\mu}\Theta-\left.\frac{\partial^{2}V}{\partial\Phi^{\dag
}\partial\Phi}\right|_{\Phi_{B}}\Theta-\left.\frac{\partial^{2}V}{\partial\Phi^{\dag
}\partial\Phi^{\dag}}\right|_{\Phi_{B}}\Theta^{\dag}=0 \, .,
\end{equation}
the last term of which, involving $\Theta^\dagger$, is not zero given the nonlinear potential chosen \eqref{eq:potential}. In fact, defining
\begin{align}
U\left(  r,\theta\right)   &  =\mu^{2}+2\lambda\phi^{2}+3\nu\phi^{4}\\
W\left(  r,\theta\right)   &  =\lambda\phi^{2}+2\nu\phi^{4}%
\end{align}
and using the ansatz for the boson star background Eq.~\eqref{eq:ansatz_phi}, we can write Eq.~\eqref{E.O.M} as
\begin{equation}
\nabla_{\mu}\nabla^{\mu}\Theta-U\left(  r,\theta\right)  \Theta-W\left(
r,\theta\right)  e^{-2i\omega_{B}t+2im_{B}\varphi}\Theta^{\dag}=0.\label{EOM}%
\end{equation}

In the following linear calculations, we assume that the contribution of $\Theta$ to the gravitational sources is negligible, fixing the background spacetime to the metric given in Eq.~\eqref{eq:LP_metric}. Under this assumption, the equation for $\Theta$ decouples from the Einstein equations, and the divergence term can be computed using the standard prescription
$\nabla_{\mu}\nabla^{\mu}\Theta  = \partial_{\mu}\left(  \sqrt{\left\vert g\right\vert }\partial^{\mu}
\Theta\right)/\sqrt{\left\vert g\right\vert }$
where the determinant of the metric in Eq.~\eqref{eq:LP_metric} is given by  $g = -A^{4}B^{2}N^{2}r^{4}\sin^{2}\theta$. We next consider the minimal case of two scattering modes, following the approach outlined in \cite{Saffin:2022tub, Zhang:2024ufh}
\begin{equation}\label{eq:ansatz_Theta}
\Theta\left(  t,\mathbf{r}\right)  =\eta^{+}\left(  r,\theta\right)
e^{-i\omega_{+}t+im_{+}\varphi}+\eta^{-}\left(  r,\theta\right)
e^{-i\omega_{-}t+im_{-}\varphi} \, .
\end{equation}
We introduce two new parameters, $\omega \in \mathbb{R}$ and $m \in \mathbb{Z}$, which are related to $\omega_\pm$ and $m_\pm$ as follows:
\begin{equation}\label{eq:omega_pm}
\omega_{\pm}=\omega_{B}\pm\omega, ~~ m_{\pm}=m_{B}\pm m \, .
\end{equation}

The general solution to Eq.~\eqref{E.O.M} can be expressed using Fourier series. The advantage of the ansatz in Eq.~\eqref{eq:ansatz_Theta} lies in its ability to factor out the dependencies on $t$ and $\varphi$, transforming the problem into a more manageable form suitable for standard 2D (linear) partial differential equation solvers. A crucial distinction from the typical boson star approach is that, while the ansatz in Eq.~\eqref{eq:ansatz_phi} is formulated to maintain stationary gravitational sources and geometry, the presence of two modes in this context enables the transfer of energy and angular momentum, as will be demonstrated in the subsequent analysis.

Substituting Eq.~\eqref{eq:ansatz_Theta} into Eq.~\eqref{EOM} and dividing by the overall factors of $\exp({-i\omega_{\pm}t+im_{\pm}\varphi})$, we derive the \textit{coupled} system of partial differential equations
\begin{equation}\label{EOM_etapm}
\begin{split}
&\frac{1}{A^{2}}\left[  \Delta_{3}\eta^\pm + \partial\ln\left(  BN\right)
\partial\eta^\pm\right]  \\
&+\left[  \left(  \frac{\omega_{\pm}-w m_{\pm}}%
{N}\right)  ^{2}-\frac{m_{\pm}^{2}}{B^{2}r^{2}\sin^{2}\theta}\right]\eta^{\pm} \\
&\qquad\qquad\quad- U\left(  r,\theta\right)  \eta^{\pm}-W\left(  r,\theta\right)
\eta^{\mp\ast}=0 \, ,
\end{split}
\end{equation}
where we have made use of the operators defined in Eq.~\eqref{eq:operators}. Now all that remains to obtain the scattering waves is to impose appropriate boundary conditions on the system ~\eqref{EOM_etapm}. It is not difficult to see that regularity in the origin implies 
\begin{equation}\label{eq:eta_at_zero}
    \eta^{\pm} \to F^{\pm}(k_{\pm} r \sin\theta)^{m_\pm} \, , ~~ r\to0
\end{equation}
with $F^{\pm}$ some dimensionless complex constants and the wave numbers,
\begin{equation}
    k_{\pm}=(\omega_\pm^2-\mu^2)^{1/2} \, ,
\end{equation}
which for propagating scattering states should be real numbers, implying the (simultaneous) condition $|\omega_{\pm}|>\mu$, or using the definition of $\omega_\pm$, equivalent to the single condition
\begin{equation}\label{eq:mass_gap}
    |\omega| > \omega_B + \mu \, .
\end{equation} 

Similarly close to the symmetry axis $\theta = 0,\pi$, the angular dependence of $\eta^{\pm}$ can be shown to be proportional to $\sin^{|m_\pm|}\theta$.

Far away from the boson star, the scalar field of the background approaches 0 exponentially and the metric approaches the Minkowski metric. In the limit $r\to\infty$, Eq.~\eqref{EOM_etapm} $\eta^+$ and $\eta^-$ decouple and the solution for the radial part, correspond to incoming and outgoing spherical waves
\begin{align}\label{eq:asymp_eta}
\eta^{\pm}  = g^{\pm}(\theta)\frac{\exp ik_{\pm}r}{k_{\pm}r}+h^{\pm}(\theta)  \frac{\exp\left(  -ik_{\pm}r\right)  }{k_{\pm}r}, ~~ r\rightarrow\infty
\end{align}
for some complex functions $g^\pm$ and $h^\pm$. This quantities can be identified, according to their group velocity of the associated waves, as ingoing and outgoing states. For concreteness:
\begin{itemize}
    \item If $\omega>0$ \qquad $\{h^+, g^-\}$ are ingoing waves and $\{g^+,h^-\}$ are outgoing waves.
    \item If $\omega<0$ \qquad $\{g^+, h^-\}$ are ingoing waves and $\{h^+,g^-\}$ are outgoing waves.
\end{itemize}

Following Ref.~\cite{Zhang:2024ufh}, we focus on the \textit{one ingoing mode}\footnote{
In Ref.~\cite{Zhang:2024ufh}, the interpretation of this particular choice is straightforward, as a partial wave decomposition is employed. Here, we opt to solve the equations directly, making the rationale behind this choice less clear. Nonetheless, the one-ingoing mode corresponds to the simplest non-trivial boundary condition (regular at the poles) that can be applied to Eq.~\eqref{EOM_etapm}.
}
which effectively selects only the + ingoing mode and adopts a simple angular dependence for this mode, specifically  $C\sin^{m_+}\theta$ , where  $C\in\mathbb{C}$  and the choice respects regularity conditions at the symmetry axis. Given the linearity of the equations, we can utilize the complex scaling $(\eta^+,\eta^-) \to (\zeta\eta^+,\zeta^*\eta^-)$ to fix the constant  $C$  in the angular dependence. This leads to the following asymptotic boundary conditions:
\begin{align}\label{eq:eta_outBC}
    \text{If }~~\omega>0, ~~g^-=0,~~h^+=\sin^{m_+}\theta \, ,\\
    \text{If }~~\omega<0, ~~h^-=0,~~g^+=\sin^{m_+}\theta \, .
\end{align}
By fixing  $C = 1$  (or any nonzero complex value), we prevent the solver from converging to the trivial solution. This approach differs from that of Ref.~\cite{Zhang:2024ufh}, where the normalization of the fields  $\eta^\pm$  is imposed at the origin.\footnote{Specifically, in \cite{Zhang:2024ufh}, the normalization condition is given by $\eta^+(0) = (k_+r\sin\theta)^{m_+}$.}

Finally, similar to the boson star background scalar field, we will focus our analysis on even-parity perturbations, although the odd-parity case is also feasible (see, e.g., Ref.~\cite{Zhang:2024ufh}). In the odd-parity case, the boundary conditions for solving the perturbations would need to be modified. For instance, instead of the condition in Eq.~\eqref{eq:eta_at_zero}, one would impose $ \eta^{\pm} \to F^{\pm}(k_{\pm} r)^{m_\pm+1}\cos\theta\sin^{m_{\pm}}\theta$ as $r \to 0$.

Given the boson star background and selected values of $\omega$ and $m$, the boundary conditions uniquely determine the solution to the linear Eq.~\eqref{EOM_etapm}.  These conditions enforce even parity, ensure regularity throughout the domain, and allow only one ingoing scattering mode. To solve this system, we employ the \texttt{Kadath} spectral library \cite{Kadath,Grandclement:2009ju}. The details about the implementation and the tests of convergence can be found in the Appendix \ref{app:tests1}. In the code the asymptotic boundary conditions \eqref{eq:eta_outBC} are enforced using the following Robin conditions for $\eta^\pm$
\begin{eqnarray}
    & r\left(k_-\eta^- - s_{\omega}i\partial_r\eta^-\right) &= 0 \, \\
    & r\left(k_+\eta^+ + s_{\omega}i\partial_r\eta^+\right) &= 2\sin^{m_+}\theta\, e^{-s_{\omega}ik_+r} ~~~~ \,
\end{eqnarray}
where $s_{\omega} = {\rm sign}(\omega)$. And they are equivalent to \eqref{eq:eta_outBC} and can be implemented in the spectral solver.

The outer boundary radius can be set as large as necessary since we interpolate the compactified background data $U$ and $W$. To prevent excessive oscillations within a single outer shell of the boson star, we maintain a fixed distance between outer shells and increase the number of shells to expand the total domain size, ensuring we are sufficiently far from the star and can achieve convergence. In the region away from the central boson star, the spacetime metric approaches the Minkowski metric. To leading order, where deviations from the Minkowski background and the sources of the $\phi$ field are negligible, the energy, momentum, and stress associated with the scattered wave can be directly computed.

In spherical coordinates, as described by Eq.~\eqref{eq:LP_metric}, and to leading order asymptotically, the vector components $n^\mu$  simplify to  $n^\mu = (1, 0, 0, 0)$, and the projection operator becomes  ${\gamma^i}_j = {\delta^i}_j$. Therefore, the source terms in Eq.~\eqref{eq:source} can be directly related to the components of the energy-momentum tensor. Specifically, the energy density is given by  $\rho = T_{tt}$ , the energy flux (momentum density) in the $r$-direction is $P^r = T^{rt}$, the angular momentum density in the  $z$-direction is  $P^\varphi = T^{\varphi t}$, and the  $z$-component of the angular momentum flux in the $r$-direction is $S_{\varphi r} = T_{\varphi r}$. Here, the energy-momentum tensor corresponds to that of a scalar field, as in Eq.~\eqref{eq:Tmunu}, but with $\Phi \to \Theta$. By inserting the asymptotic behavior from Eq.~\eqref{eq:asymp_eta}, we obtain the following leading-order expressions:
\begin{align}
    \rho & =  \left |  \partial_t \Theta \right |^2 + \left | \partial_r \Theta \right |^2 + \mu^2\left | \Theta \right |^2            \, , \\
    P^r  & =  -\partial_r \Theta^* \partial_t \Theta - \partial_r \Theta \partial_t \Theta^*      \, , \\
    P^\varphi & =  -\partial_t \Theta^* \partial_\varphi \Theta - \partial_t \Theta \partial_\varphi \Theta^*    \, , \\
    S_{\varphi r} & = \partial_r \Theta^* \partial_\varphi \Theta + \partial_r \Theta \partial_\varphi \Theta^* \, .
\end{align}

To determine the energy and angular momentum enhancement of asymptotic waves scattering off a boson star, we introduce amplification factors. Following previous works \cite{Saffin:2022tub,Cardoso:2023dtm,Gao:2023gof,Zhang:2024ufh}, there are two common ways to define these amplification factors. 

The energy and angular momentum densities in an asymptotic spherical shell, bounded by an interior radius $r_1$ and an exterior radius $r_2$, are given by expressions that simplify further for large radii. These reduced forms capture the dominant behavior of the energy and angular momentum as waves propagate through the asymptotic region:
\begin{align}
    E_{\circledcirc} &= \frac{1}{r_2-r_1}\int_{r_1}^{r_2}r^2\mathrm{d}r \left \langle \rho \right \rangle_{T\Omega} \nonumber\\
                     &\propto \frac{\omega_+^2}{k_+^2}\left ( \left | A_+ \right |^2 + \left | B_+ \right |^2 \right ) + \frac{\omega_-^2}{k_-^2}\left ( \left | A_- \right |^2 + \left | B_- \right |^2 \right )\\
    L_{\circledcirc} &= \frac{1}{r_2-r_1}\int_{r_1}^{r_2}r^2\mathrm{d}r \left \langle P^{\varphi} \right \rangle_{T\Omega} \nonumber\\
                     &\propto  \frac{\omega_+m_+}{k_+} \left( \left | A_+ \right |^2 + \left | B_+ \right |^2 \right ) + \frac{\omega_-m_-}{k_-} \left( \left | A_- \right |^2 + \left | B_- \right |^2 \right ) 
\end{align}
where $r_2-r_1$ includes at least a full spatial oscillation of the longest wave, $\left \langle  \right \rangle_{T\Omega} $ denotes time average over a few oscillations and solid angle average over the 2-sphere. Also here we used the definitions
\begin{subequations}\label{eq:A_and_B}
\begin{align}
    \left |  A_\pm \right |^2 &= \int_{0}^\pi \mathrm{d}\theta \sin\theta \left | g^{\pm}(\theta) \right |^2 \, ,  \\
    \left |  B_\pm \right |^2 &= \int_{0}^\pi \mathrm{d}\theta \sin\theta \left | h^{\pm}(\theta) \right |^2  \, .
\end{align}
\end{subequations}
So, the first kind of amplification factor can be defined as the ratio of energy/angular momentum in the outgoing mode, compared to the ingoing one,
\begin{align}
    \mathcal{A}_{E} &= \left ( \frac{\frac{\omega_+^2}{k_+^2}\left| A_+ \right|^2+\frac{\omega_-^2}{k_-^2}\left| B_- \right|^2}{\frac{\omega_-^2}{k_-^2}\left| A_- \right|^2+\frac{\omega_+^2}{k_+^2}\left| B_+ \right|^2} \right ) ^{\mathrm{sign}(\omega)} \,, \\
    \mathcal{A}_{L} &= \left ( \frac{\frac{\omega_+m_+}{k_+^2}\left| A_+ \right|^2+\frac{\omega_-m_-}{k_-^2}\left| B_- \right|^2}{\frac{\omega_-m_-}{k_-^2}\left| A_- \right|^2+\frac{\omega_+m_+}{k_+^2}\left| B_+ \right|^2} \right )^{\mathrm{sign}(\omega)} \, . \label{eq:AL}
\end{align}

The second way of calculating the amplification factors, pointed out in ~\cite{Cardoso:2023dtm}, is to utilise fluxes at spatial infinity:
\begin{align}\label{eq:fluxes}
    \mathcal{F}_{E}^{\infty} &= \lim_{r\rightarrow \infty}r^2\left \langle P^r \right \rangle_{T\Omega} \nonumber\\
                             &\propto \frac{\omega_+}{k_+}\left ( \left | A_+ \right |^2-\left | B_+ \right |^2  \right )+\frac{\omega_-}{k_-}\left ( \left | A_- \right |^2-\left | B_- \right |^2  \right ) \, ,\\
    \mathcal{F}_{L}^{\infty} &= \lim_{r\rightarrow \infty} r^2\left \langle S^{r\varphi} \right \rangle_{T\Omega} \nonumber\\
                             &\propto \frac{m_+}{k_+}\left ( \left | A_+ \right |^2-\left | B_+ \right |^2  \right )+\frac{m_-}{k_-}\left ( \left | A_- \right |^2-\left | B_- \right |^2  \right ) \, .\label{eq:fluxesL}
\end{align}
Similarly,  we can also define the energy/angular momentum flux amplification factors. For $\omega > 0$, the ingoing energy flux corresponds to the (negative) contributions of the ingoing modes $A_-$ and $B_+$ in Eq.~\eqref{eq:fluxes}, while the outgoing energy flux arises from the contributions of the outgoing modes $A_+$ and $B_-$. A similar reasoning applies for the case $\omega < 0$, enabling the calculation of the energy flux amplification factor $\mathcal{A}_{tr}$. By following the same approach and using Eq.~\eqref{eq:fluxesL}, the angular momentum flux amplification factor $\mathcal{A}_{r\varphi}$ can also be determined, resulting in the following expressions:
\begin{align}
    \mathcal{A}_{tr} &= \left ( \frac{\frac{\omega_+}{k_+}\left| A_+ \right|^2-\frac{\omega_-}{k_-}\left| B_- \right|^2}{\frac{\omega_+}{k_+}\left| B_+ \right|^2-\frac{\omega_-}{k_-}\left| A_- \right|^2} \right ) ^{\mathrm{sign}(\omega)} \, , \label{eq:Atr}\\
    \mathcal{A}_{r\varphi} &= \left ( \frac{\frac{m_+}{k_+}\left| A_+ \right|^2-\frac{m_-}{k_-}\left| B_- \right|^2}{\frac{m_+}{k_+}\left| B_+ \right|^2-\frac{m_-}{k_-}\left| A_- \right|^2} \right )^{\mathrm{sign}(\omega)} \, . \label{eq:Arphi}
\end{align}
For a monochromatic wave, the definitions of energy amplification, $\mathcal{A}_E$ and $\mathcal{A}_{tr}$, coincide, as do the corresponding angular momentum amplification factors, $\mathcal{A}_L$ and $\mathcal{A}_{r\varphi}$.

Next, we highlight a key feature of the $U(1)$ symmetry in the perturbative field inherited from the $U(1)$ symmetry of the full field. This symmetry guarantees the conservation of a ``particle number'' during the scattering process. The amplification factor is given by (see, e.g., \cite{Gao:2023gof, Saffin:2022tub, Zhang:2024ufh}):

\begin{equation}\label{eq:AN}
    \mathcal{A}_N=\left ( \frac{\frac{1}{k_+}\left| A_+ \right|^2+\frac{1}{k_-}\left| B_- \right|^2}{\frac{1}{k_-}\left| A_- \right|^2+\frac{1}{k_+}\left| B_+ \right|^2} \right ) ^{\mathrm{sign}(\omega)}.
\end{equation}
The conservation of particle number means $\mathcal{A}_N=1$, i.e.
\begin{equation}\label{eq:conservation}
    \frac{1}{k_+}\left| A_+ \right|^2+\frac{1}{k_-}\left| B_- \right|^2 = \frac{1}{k_-}\left| A_- \right|^2+\frac{1}{k_+}\left| B_+ \right|^2
\end{equation}
This property can be used to derive some superradiance criteria. Furthermore, it serves to verify the consistency of our numerical results.

We now describe the process by which we extract the four quantities $|A_{\pm}|^2$, $|B_{\pm}|^2$ from the solutions $\eta_+(r,\theta)$ and $\eta_-(r,\theta)$. To achieve this, we employed two distinct methods for calculating these quantities and used their relative difference as an error indicator in our code, alongside the constant $\mathcal{A}_N$.

In the first method, we formed combinations of $\eta_{\pm}$ and $\partial_r \eta_{\pm}$ to express the real and imaginary parts of the functions $g^{\pm}(\theta)$ and $h^{\pm}(\theta)$ using Eq.~\eqref{eq:asymp_eta}. We evaluated these expressions at the outer boundary and then substituted the results into the surface integrals given in Eq.~\eqref{eq:A_and_B}. The second method involved using specific volume integrals within a spherical shell, defined by an interior radius $\mu r_1 \gg 1$ and an exterior radius $r_2 > r_1$. In this approach, we integrated the quantities $\eta^{\pm*} \partial_r \eta^\pm$ and $\eta^\pm \eta^{\pm*}$, and subsequently formed linear combinations of the resulting four real numbers to reconstruct the coefficients $|A_{\pm}|^2$ and $|B_{\pm}|^2$.
With this information, along with the necessary tests establishing the precision of the code, we proceeded to calculate the amplification factors for specific cases of interest.

\begin{figure}
\centering
    \includegraphics[width=0.48\textwidth]{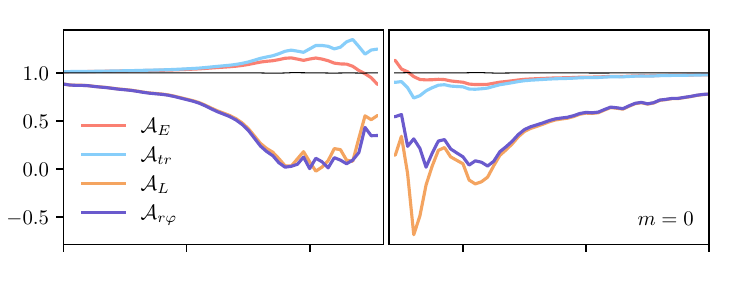}\\\vspace{-0.6cm}
    \includegraphics[width=0.48\textwidth]{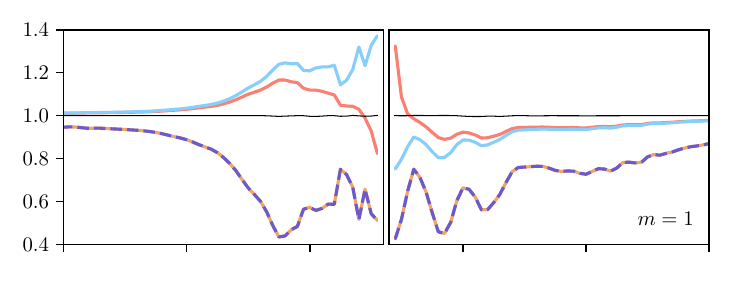}\\\vspace{-0.6cm}
    \includegraphics[width=0.48\textwidth]{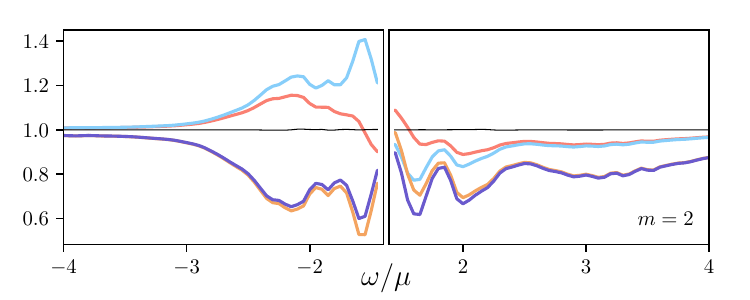} 
  \caption{
  Amplification factors as a function of the frequency parameter $\omega$ for three different values of $m$. The background configuration correspond to a spinning boson star with $m_B = 1$ and $\omega_B = 0.4\mu$. This is one of the fiducial configurations present in panel (b) of Fig.~\ref{fig:M-spherical}. The thin black line correspond to the quantity $\mathcal{A}_N$ which is consistent with 1 in all the cases. The empty gap in $\omega$ corresponds to the mass gap \eqref{eq:mass_gap}, which in this case is the interval $\omega\in(-1.4,1.4)\mu$.
  }
  \label{fig:factors_0p4}
\end{figure}
\begin{figure}
\centering
    \includegraphics[width=0.48\textwidth]{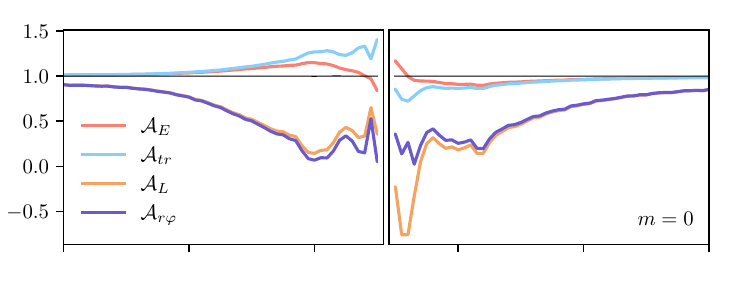}\\\vspace{-0.6cm}
    \includegraphics[width=0.48\textwidth]{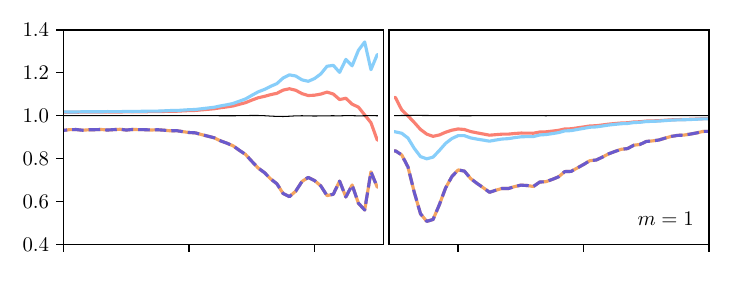}\\\vspace{-0.6cm}
    \includegraphics[width=0.48\textwidth]{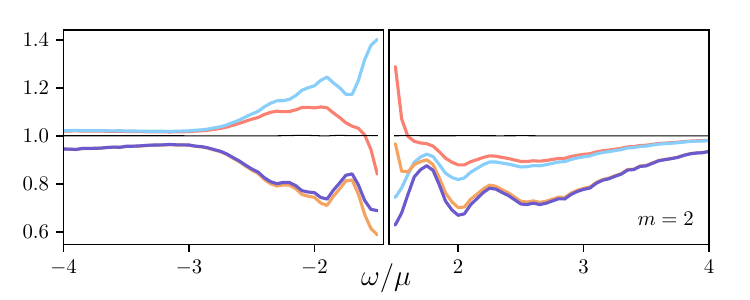} 
  \caption{
  Same as Fig.~\ref{fig:factors_0p4} but with a background solution $\omega_B=0.45\mu$, also a fiducial configuration shown in Fig.~\ref{fig:M-spherical}. The mass gap correspond to the interval $\omega\in(-1.45,1.45)\mu$.
  }
  \label{fig:factors_0p45}
\end{figure}

Since the aim of this paper is to investigate not only the linear amplification of waves around a boson star but also the dynamics of a wavepacket scattering off a boson star through nonlinear simulations. In this context, the stable star configurations are of primary importance for our analysis. To this end, we conducted stability checks on several solutions, identifying stable cases in both spherical and rotating scenarios. Specifically, for the rotating case, we selected models with $\sigma = 0.05$ (where $g = -127.3$ and $h = 3039.6$) and frequencies $\omega_B = 0.4$ and 0.45. These models correspond to the purple diamonds depicted in the (b) subplot of Fig. \ref{fig:M-spherical}.

Using this pair of rotating configurations, we calculate the amplification factors and present their spectra in Fig.~\ref{fig:factors_0p4} for the case $m_B = 1$ and $\omega_B = 0.4\mu$, and in Fig.~\ref{fig:factors_0p45} for $m_B = 1$ and $\omega_B = 0.45\mu$. These calculations consider waves with azimuthal parameter $m = 0$, 1, 2, corresponding to $m_+ = 1$, 2, 3, and show the amplification factor in regions where waves propagate as defined by Eq.~\eqref{eq:mass_gap}. Additional details on the implementation and verification of these results are provided in Appendix~\ref{app:tests}. Here, we note that the sharp features observed in some of the amplification factors are intrinsic to the solution and remain present when the resolution is increased or additional data points are included. Specifically, for the results presented, the step size for $\omega$ used by the solver to generate the amplification factor plots is $0.05$.

From Figs.~\ref{fig:factors_0p4} and \ref{fig:factors_0p45}, we observe that, far from the mass gap, the amplification factors associated with energy, $\mathcal{A}_E$ and $\mathcal{A}_{tr}$, and those associated with angular momentum, $\mathcal{A}_L$ and $\mathcal{A}_{r\varphi}$, converge to similar values. This convergence occurs because the differences between the group velocities of the + and $-$ modes (arising from the mass term in the Klein-Gordon field) diminish as $|\omega| \to \infty$ \cite{Saffin:2022tub}. Specifically, in this limit, all four quantities, $\omega_+$, $-\omega_-$, $k_+$, and $k_-$, tend toward $\omega$. In contrast, near the mass gap $|\omega| = \omega_B + \mu$, significant differences emerge. The amplification factors associated with densities, $\mathcal{A}_E$ and $\mathcal{A}_L$, represent the enhancement of energy or angular momentum in the outer region after the interaction. In contrast, the flux-related amplification factors, $\mathcal{A}_{tr}$ and $\mathcal{A}_{r\varphi}$, quantify (at the linear level) the extraction of energy or angular momentum from the central body. For example, near the mass gap and for $\omega > 0$, we observe suppressed flux but amplified densities, indicating greater accumulation of energy in the distant region after passing through the star. These interpretations are validated in Sec.~\ref{sec:nonlinear}, where the dynamics of wavepackets are analyzed.

As $|\omega| \to \infty$, we expect all amplification factors to approach 1. For $\omega \to \infty$, $\mathcal{A}_E$, $\mathcal{A}_{tr}$, $\mathcal{A}_L$, and $\mathcal{A}_{r\varphi}$ approach 1 from below. However, for $\omega \to -\infty$, $\mathcal{A}_E$ and $\mathcal{A}_{tr}$ approach 1 from above, while $\mathcal{A}_L$ and $\mathcal{A}_{r\varphi}$ approach from below. Additionally, for $m = 1$ waves, the two angular momentum amplification factors, $\mathcal{A}_L$ and $\mathcal{A}_{r\varphi}$, coincide, causing their corresponding lines to overlap. This can be verified by noting that $m_- = 0$, which implies the equivalence of Eqs.~\eqref{eq:AL} and \eqref{eq:Arphi}, leading to $\mathcal{A}_L = \mathcal{A}_{r\varphi}$.

From the figures, we observe that, consistent with previous studies \cite{Cardoso:2023dtm,Gao:2023gof,Zhang:2024ufh}, the energy flux amplification factor $\mathcal{A}_{tr}$ follows a Zeldovich-like superradiance condition. Specifically, $\mathcal{A}_{tr} > 1$ when $\omega < -(1 + \omega_B)$ (or equivalently, $\omega_+ < \omega_B$). This condition can be explicitly verified by combining the amplification factor definition \eqref{eq:Atr} and the particle number conservation law \eqref{eq:conservation}, which yields:
\begin{equation}
    \mathcal{A}_{tr} = \frac{p\omega_+ /k_+ - \omega_-/k_-}{p\omega_+/k_++\omega_+/k_-} >1 \, ,
\end{equation}
where $p = |A_+|^2 / |B_-|^2$ for positive $\omega$ and $p = |B_+|^2 / |A_-|^2$ for negative $\omega$ with only one ingoing mode. Since $p$ is a non-negative quantity, the inequality simplifies to the expected condition $\omega < -(1 + \omega_B)$. Within this frequency range, the amplification factor is bounded above by $-\omega_-/\omega_+$ \cite{Zhang:2024ufh}. 

A similar Zeldovich-like condition governs the amplification of angular momentum flux. In this case, using Eqs.~\eqref{eq:Arphi} and \eqref{eq:conservation}, the amplification factor $\mathcal{A}_{r\varphi}$ is given by:
\begin{equation}
    \mathcal{A}_{r\varphi} = \frac{p-m_-k_+/(m_+k_-)}{p+k_+/k_-} >1 \, ,
\end{equation}
which implies the condition $m < -m_B$. This result aligns with the cases explored in Figs.~\ref{fig:factors_0p4} and \ref{fig:factors_0p45}. Notably, the angular momentum flux can also be amplified in the opposite direction ($\mathcal{A}_{r\varphi} < -1$), but this scenario is subject to a necessary, though not sufficient, condition $-m_B<m<0$.

The classical Zeldovich rotational superradiance condition states that the incident wave’s frequency, such as $\omega_+$ for the $\eta_+$ wave, must be smaller than the product of the rotating body’s angular frequency, $\Omega_B$, and the wave’s azimuthal number, $m_+$. For a rotating boson star, the angular frequency is defined as $\Omega_B = \omega_B / m_B$. However, unlike systems with a dissipative mechanism (e.g., a rotating cylinder) or a one-way membrane (e.g., a black hole’s event horizon), rotating boson stars lack a mechanism to extract negative energy modes. Consequently, the condition $\mathcal{A}_{r\varphi} > 1$ which requires $m_+ < 0$, differs from the Zeldovich condition $m_+ > m_B \omega_+ / \omega_B$.

Finally we verify that the amplification of energy and angular momentum densities are consistent with the analytical bounds \cite{Saffin:2022tub,Zhang:2024ufh},
\begin{equation}
    1<\mathcal{A}_E<\frac{\omega_-^2 k_+}{\omega_+^2 k_-} \, ,
\end{equation}
for $\omega<-\omega_E$ or $\omega_B+\mu<\omega<\omega_E$ with
\begin{equation}
    \omega_E = \left[\mu^2+\omega_B^2 + \left(\mu^2+4\omega_B^2\right)^{1/2} \right]^{1/2} \, .
\end{equation}
And
\begin{equation}
    1<\mathcal{A}_L<\frac{\omega_-m_- k_+}{\omega_+m_+ k_-} \, ,
\end{equation}
for the frequency region $\mu+\omega_B<\omega<\omega_L$ if $m>m_B$ and the regions $\omega<\omega_L$, $\mu+\omega_B<\omega$ if $m<-m_B$. Here, the critical frequency $\omega_L$ satisfy the condition,
\begin{equation}
    \frac{(\omega_B+\omega_L)(\omega_B+m_L)}{\sqrt{(\omega_B+\omega_L)^2-\mu^2}} = \frac{(\omega_B-\omega_L)(\omega_B-m_L)}{\sqrt{(\omega_B-\omega_L)^2-\mu^2}} \, .
\end{equation}
We have found that the obtained results fall within the previous bounds.


\section{Nonlinear wavepacket scattering}\label{sec:nonlinear}

Alternatively to the linear analysis presented in Sec.~\ref{sec:linear}, the potential amplification of scalar waves incident on a star can be dynamically analyzed through evolutions. First, we examine waves with low amplitudes to verify the previously obtained amplification factors; then, we explore the validity range of the linear approximation and study the nonlinear effects arising from the star-wave interaction and backreaction on the spacetime. To this end, we first outline how the wave signal is prepared on a boson star spacetime background, then describe the specific implementation and results for spherical waves in a spherical star background and the general scenario of a non-spherical wave on a star background, whether rotating or not. Finally we consider the case of spherical waves in a cavity with a boson star inside.

\subsection{Initial data}\label{sec:nonlinear_id}

We begin by considering a spherically symmetric implementation, setting $m_B = 0$ in the ansatz for the background boson star scalar field, Eq.~\eqref{eq:ansatz_phi}, and assuming $\Psi^2 := e^{F_1} = e^{F_2}$ in the background metric ansatz, Eq.~\eqref{eq:LP_metric}. We further restrict our analysis to a spherically symmetric wavepacket ($m_\pm = 0$). Here, we will work in isotropic coordinates and refer to $\Psi$ as the conformal factor. In this case, the line element simplifies to:
\begin{equation}\label{eq:BS_metric}
ds^2=-\alpha^2(r)dt^2+\Psi^4(r)\left(dr^2+r^2d\Omega^2\right) \, .
\end{equation}
In what follows, we denote the boson star solution as $\alpha_{\rm B}$ and $\Psi_{\rm B}$, representing any solution of Eq.~\eqref{eq:BS_metric} obtained by solving the Einstein and Klein-Gordon equations with the corresponding stationary ansatz.

Now, we introduce an incoming wavepacket outside the boson star, centered around $r = r_0$, with a frequency $\omega_0$ and a width $\sigma_r \ll r_0$, at the initial time,
\begin{equation}\label{eq:Theta}
\begin{split}
  \Theta(r) =& \frac{\delta\, \phi(0)}{r_0} \exp\left[-\frac{(r-r_0)^2}{2\sigma_r^2}\right] \\
  &\times \exp\left[-i s_{\omega_0}\sqrt{\omega_0^2-1}\, r - i\omega_0 t\right]_{t=0} \, ,
\end{split}
\end{equation}
$\phi(0)$ is the boson star scalar field profile \eqref{eq:ansatz_phi} evaluated at $r=0$, $\delta$ some number to control the amplitude of the incoming wavepacket and $\omega_0=\omega_{\rm B}+\omega$. The quantity $\omega_0$ coincides with $\omega_+$, defined in Eq.~\eqref{eq:omega_pm}. This is because we want to consider the wavepacket as the $+$ ingoing mode of the amplification factors calculated in Sec.~\ref{sec:linear}. After the wavepacket hits the boson star we expect to have the two outgoing modes with frequencies $\omega_+$ and $\omega_-$, so the simultaneous condition $|\omega_\pm|>\mu$, discussed in Sec.~\ref{sec:linear}  needs to be satisfied, implying that we should prepare the wavepacket such that
\begin{equation}
    |\omega_0|>\mu \,~~ \mathrm{and} ~~|2\omega_B-\omega_0|>\mu \, .
\end{equation}

Next, we combine both lumps of scalar field at $t=0$,
\begin{equation}\label{eq:fullPhi}
\Phi = \Phi_{\rm B} + \Theta \, ,
\end{equation}
Clearly, whenever $\delta\neq0$ the metric~\eqref{eq:BS_metric} is no longer 
a solution to the Einstein equations. To correct this, we must re-solve for 
the geometrical quantities.

In spherical symmetry the 3+1 line element
becomes (see Sec.~\ref{sec:evol_scheme_sph} or for instance ~\cite{Alcubierre:2011pkc}),
\begin{equation}\label{eq:metric_spherical}
  ds^2=(-\alpha^2+a\beta^2)dt^2+2a\beta dt dr + a dr^2 + b r^2d\Omega^2 \, ,
\end{equation}
and the extrinsic curvature has only two independent components; 
say $K$ and one of the non-zero components of the traceless part of the conformal
extrinsic curvature $\tilde{A}_{ij}=K_{ij}-\frac{1}{3}\gamma_{ij}K$.
Applying these ansatz in the constraints 
Eqs.~\eqref{eq:Hamiltonian} and \eqref{eq:momentumConstraint} the following expressions are obtained:
\begin{eqnarray}
  && {}^{(3)}R-(A^{2}_{a}+2A_{b}^{2})+\frac{2}{3}K^{2}-16\pi \rho=0,\label{eq:hamiltonian}\\
  &&\partial_{r}A_{a}-\frac{2}{3}\partial_{r}K+(A_{a}-A_{b})\left(\frac{2}{r}+\frac{\partial_{r}b}{b}\right)-8\pi P_{r}=0. ~~~~~~\label{eq:momentum}
\end{eqnarray}
Here, $A_a:=\tilde{A}^r_{\;\;r}$ and $A_b:=\tilde{A}^\theta_{\;\;\theta}=\tilde{A}^\varphi_{\;\;\varphi}$, 
with $A_a+2A_b=0$ by definition. When $\delta=0$ it is clear that $P_r=0$ (the isolated boson star is a static solution, but of course it can also be checked directly). 

Now, we choose to solve for the initial data in isotropic coordinates, 
\begin{equation}
    a=b=\Psi^4 \, .
\end{equation}
The metric coefficients will depart from the boson star solution, with conformal factor $\Psi$, as $\delta$ increases. We arrive at the following equations for the functions $A_a$ and the updated function $\Psi$, using Eq.~\eqref{eq:momentum} and Eq.~\eqref{eq:hamiltonian}, respectively:
\begin{eqnarray}
  &&\p_r A_a + \frac{3}{r}A_a+6A_a\frac{\p_r\Psi}{\Psi} - 8\pi P_r = 0\, ,\\
  &&-\frac{8}{\Psi^5}\left(\partial_{r}^2\Psi+\frac{2}{r}\partial_r\Psi\right)-\frac{3}{2}A_a^2-16\pi\rho=0 \,\label{eq:newPsi} ;
\end{eqnarray}
where $\rho$ and $P_r$ are obtained from the full $\Phi$ in Eq.~\eqref{eq:fullPhi}, obtaining
\begin{eqnarray}
  \rho & = & \left(\frac{1}{\Psi^4}+\frac{\beta^2}{\alpha^2}\right)\partial_r\Phi \partial_r\Phi^* + V \nonumber\\
  &&+ \frac{1}{\alpha^2}\left( \partial_{t}\Phi\partial_{t}\Phi^{\ast}-\beta
\partial_{t}\Phi\partial_{r}\Phi^{\ast}-\beta\partial_{t}\Phi^{\ast}
\partial_{r}\Phi \right) \, \label{eq:rho},~~\\
  P_r & = & -\frac{1}{\alpha}(\partial_r\Phi \partial_t\Phi^* + \partial_r\Phi^* \partial_t\Phi -2\beta\partial_{r}\Phi^{\ast}\partial_{r}\Phi )\, \label{eq:j_r}. 
\end{eqnarray}

We choose $\alpha=\alpha_{\rm B}$ and $\beta = 0$, which simplifies the construction of initial data as well as being consistent with the numerical evolution code (which uses the 1+log condition for the lapse and zero shift for the spherical evolution). Furthermore, the expressions for the matter sources \eqref{eq:rho}-\eqref{eq:j_r} can be approximated at $t=0$ using the fact the both lumps of scalar field are far apart from each other, $r_0\gg R_{99}$.
In this case we have 
\begin{align}
\rho(t=0)  \simeq \rho_{B} + \rho_{\Theta} \, , ~~ P_r(t=0)  \simeq  P_{\Theta} \, ,
\end{align}
with
\begin{align}
P_{\Theta}   &  :=-\frac{1}{\alpha}\left(  \partial
_{r}\Theta\partial_{t}\Theta^{\ast}+\partial_{r}\Theta^{\ast}\partial
_{t}\Theta\right)  \nonumber\\
&  =-\tilde{\delta}^2 \frac{2\omega_{0}s_{\omega_{0}}\sqrt{\omega_{0}^{2}-1}}{\alpha}%
\exp\left[  -\frac{\left(  r-r_{0}\right)  ^{2}}{\sigma_r^{2}}\right] \, ,
\end{align}
\begin{align}
\rho_{\Theta}   &  :=  \frac{1}{\Psi^{4}}\partial_{r}%
\Theta\partial_{r}\Theta^{\ast}+\frac{1}{\alpha^{2}}\partial_{t}\Theta
\partial_{t}\Theta^{\ast}+V(|\Theta|) \nonumber\\
&  = \frac{\tilde{\delta}^2}{\Psi^{4}}\left[  \left(  \frac{r-r_{0}}{\sigma_r^{2}}\right)
^{2}+\omega_{0}^{2}-1 + \omega_0^2\frac{\Psi^4}{\alpha^2}\right]\nonumber\\
&\quad\times \exp\left[  -\frac{\left(  r-r_{0}\right)  ^{2}%
}{\sigma_r^{2}}\right]   +V(|\Theta|)\, ,
\end{align}
and $\tilde{\delta} = \delta\, \phi(0)/r_0$.

This concludes the necessary details for setting the initial data in the spherical case, which will be used in the evolution code described in Section \ref{sec:evol_scheme_sph}. We now turn to the case of generic $m_B$ and $m_\pm$, which involves a different implementation and presents its own challenges. In this scenario, we do not solve the constraint equations after adding the wavepacket. Therefore, in our evolutions, we limit the analysis to cases where the wavepacket’s amplitude is sufficiently small, ensuring that the induced errors remain controlled.

Analogous to the spherical wavepacket defined in Eq.~\eqref{eq:Theta}, we introduce the parameter $m_0$, which plays a similar role to that of $m_+$ in the $+$ mode of the perturbative wave:
\begin{equation}\label{eq:Theta_3D}
\begin{split}
    \Theta(\mathbf{r}) = &\frac{\delta\, \phi_{\rm max}}{r_0} \sin^{m_0}\theta\, \exp\left[-\frac{(r-r_0)^2}{2\sigma_r^2}\right]\, \\
  \times&\exp\left[-i s_{\omega_0}\sqrt{\omega_0^2-1}\, r - i\omega_0 t + i m_0 \varphi\right]_{t=0} \, ,
\end{split}
\end{equation}
where $\phi_{\rm max}$ represent the absolute maximum of the boson star field $\phi$ (which for $m_B>0$ is not at $r=0$). Then we add the boson star background and the wavepacket as in Eq.~\eqref{eq:fullPhi} and fix the metric coefficients $F_0$, $F_1$, $F_2$ and $w$ to those of the boson star in Lewis-Papapetrou coordinates \eqref{eq:LP_metric}. Then, in order to set initial data in the 3D implementation presented in Sec.~\ref{sec:evol_scheme_3D} we have to transform all the objects to Cartesian coordinates. 

We transform initial data generated in spherical coordinates $x^{i'} = (r,\theta,\varphi)$ to the Cartesian coordinates $x^i=(x,y,z)$, used in the evolution. We notice that the metric in Eq.~\eqref{eq:LP_metric} is related to the 3+1 line element Eq.~\eqref{eq:3+1_metric} through $\alpha = N$, $\gamma_{i'j'}={\rm diag}(A^2,A^2r^2,B^2r^2\sin^2\theta)$ and $\beta^{i'} = (0,0,-w)$. Then, the components of the intrinsic metric and shift vector in the Cartesian coordinate base can be calculated using the Jacobian matrices $\partial x^{i'}/\partial {x^j}$ and $\partial {x^i}/\partial x^{j'}$. In particular we obtain that the shift vector at $t=0$ is given by $\beta^i = (yw,-xw,0)$. We need to provide $K_{ij}$ at $t=0$ to the evolution code, for this we first notice that the only nonzero components in spherical coordinates are $K_{A\varphi} = K_{\varphi A} = -B^2 r^2\sin^2\theta \partial_A w /(2N)$, with $A=r$ or $\theta$. And after this we transform $K_{ij}$ as $\gamma_{ij}$ using the Jacobian matrices.

We then interpolate the 2D data to the 3D infrastructure of the \texttt{Einstein Toolkit} using the procedures and codes detailed in \cite{Jaramillo:2024smx}. For the scalar field, interpolation is the only necessary step. Utilizing Eq.~\eqref{eq:Kphi}, the ansatz \eqref{eq:ansatz_phi}, the perturbation \eqref{eq:Theta_3D}, and the fact that $\Lie_\beta\Phi = \beta^i\partial_i\Phi = -w\partial_\varphi\Phi$, we obtain the initial value for the time derivative of the scalar field, $K_\Phi$, at $t = 0$ as:
\begin{equation}
    K_\Phi (t=0, \mathbf{r}) = \frac{i}{2\alpha}\left[(\omega_B-m_Bw)\Phi_B + (\omega_0 - m_0 w)\Theta \right] \, .
\end{equation}

Having outlined the methodology for setting up the initial data, we now turn to presenting the specific cases we analyzed and the corresponding results. The initial configurations were carefully selected based on their physical relevance and potential to highlight key dynamical features. In particular, we examined both spherical and rotating boson stars, with varying frequencies and wavepacket parameters to explore a wide range of scattering phenomena.

For each case, we systematically varied the wavepacket characteristics, such as the central frequency $\omega_0$ and angular momentum modes $m_\pm$, to study how these factors influence the amplification, attenuation, and overall dynamics of the system. In the following sections, we will detail these cases individually, discussing both the setup and the key outcomes of each simulation.

\subsection{Spherical symmetry}

We perform spherically symmetric evolutions based on the two fiducial boson star configurations shown in Fig.~\ref{fig:M-spherical}, panel (a). These configurations are stable, characterized by $g = -10= -h$ and the frequencies $\omega_B = 0.5\mu$ and $\omega_B = 0.6\mu$. Additionally, we consider the mini-boson star case ($g = 0$ and $h = 0$), and a frequency of $\omega_B = 0.9\mu$, which is also known to be stable \cite{Gleiser:1988ih,Seidel:1990jh}.

In the case of a free scalar field potential, the modes $\eta^{\pm}$ in the linear perturbation equation \eqref{EOM_etapm} decouple, resulting in no amplification or attenuation of the perturbations, as previously noted in \cite{Saffin:2022tub}. Thus, we use the free-field mini-boson star as a control case to confirm the validity of the linear regime in this model.

For these three background solutions, we introduce the wavepacket from Eq.~\eqref{eq:Theta}, which requires selecting values for the four free parameters of the wavepacket: $\delta$, $\omega_0$, $\sigma_r$, and $r_0$. We set the initial position parameter, $r_0$, sufficiently far from the boson star so that an extraction radius, $r_e$, can be placed outside the star but still within the “inner part” of the wavepacket. Specifically, we ensure the following condition is met:
\begin{equation}
    R_{99} < r_e \lesssim r_0 - \sigma_r \, .
\end{equation}
At the extraction surface $r = r_e$, we will calculate various quantities to determine the amplification factors. For the wavepacket’s width, $\sigma_r$, we select a value large enough that for a given range of $\omega_0$, a few oscillations are clearly visible within the interval $r \in (r_0 - \sigma_r, r_0 + \sigma_r)$. Thus, with careful selection of both parameters $r_0$ and $\sigma_r$, they are not expected to play a critical role in the scattering process of the wavepacket, allowing us to focus on the physical interaction with the boson star.

On the other hand, the parameters $\omega_0$ and $\delta$ play a central role in this analysis. As previously discussed, the domain of interest for $\omega_0$ is defined by the conditions $|\omega_0| > \mu$ and $|2\omega_B - \omega_0| > \mu$. Regarding the amplitude $\delta$, we anticipate that in the limit $\delta \to 0$, we should recover the results from the linear regime. However, for higher values of $\delta$, significant deviations from the linear results presented in Sec.~\ref{sec:linear} can emerge, potentially leading to gravitational collapse in certain cases, as we will explore. We begin by systematically analyzing the effects of varying $\delta$ and $\omega_0$. Table~\ref{tab:spherical_cases} summarizes the parameters for the background and wavepacket used in this exploration, along with some data on the global properties of the boson star.
\begin{table}[b]
    \begin{tabular}{cc|cccc|cccc}
    \hline
        $g$    & $h$ & $\omega_B/\mu$ & $M\mu$ & $Q\mu^2$ & $R_{99}\mu$ & $\log_{10}\delta$ & $\omega/\mu$ & $r_0\mu$ & $\sigma_r\mu$ \\ \hline
        $-10$  & 10  & 0.6      & 0.345  & 0.357    & 3.67        & $[{-3},{-1}]$ & $\pm2,\pm3$ & 40 & 5\\
        $-10$  & 10  & 0.5      & 0.392  & 0.442    & 3.51        & $[{-4},{-2}]$ & $\pm2,\pm3$ & 40 & 5\\
        $0$    & 0   & 0.9      & 0.605  & 0.621    & 10.5        & $[{-3},0]$       & $\pm2,\pm3$ & 40 & 5\\
    \hline
    \end{tabular}
    \caption{Configurations used in the scan over the parameter space for a spherically symmetric wavepacket \eqref{eq:Theta} in a static boson star background. Instead of $\omega_0$, we report the quantity $\omega = \omega_0 - \omega_B$ for convenience in the analysis below. For the amplitude $\delta$, the ranges comprise 5 to 11 points.}
    \label{tab:spherical_cases}
\end{table}

\begin{figure}
    \centering
    \subfigure[~~mini-boson star]{
       \hspace{0.6cm}\includegraphics[width=0.45\textwidth]{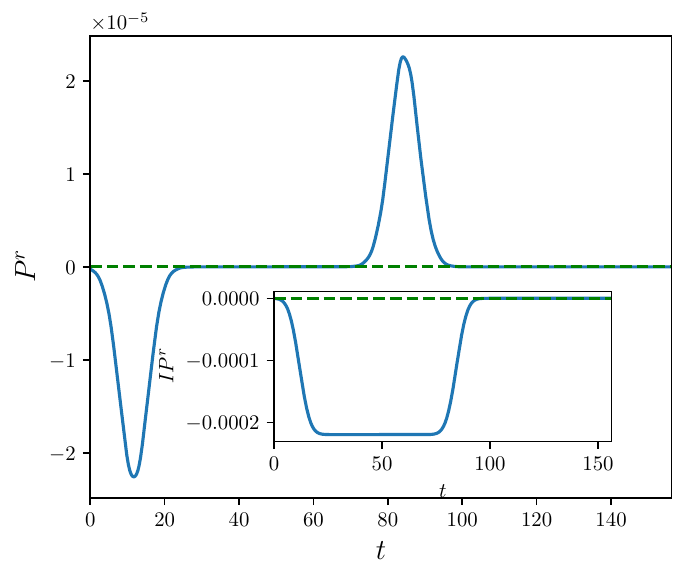}
      }
    \subfigure[~~boson star with $g=-10,h=10,\omega_B=0.6\mu$]{
       \includegraphics[width=0.48\textwidth]{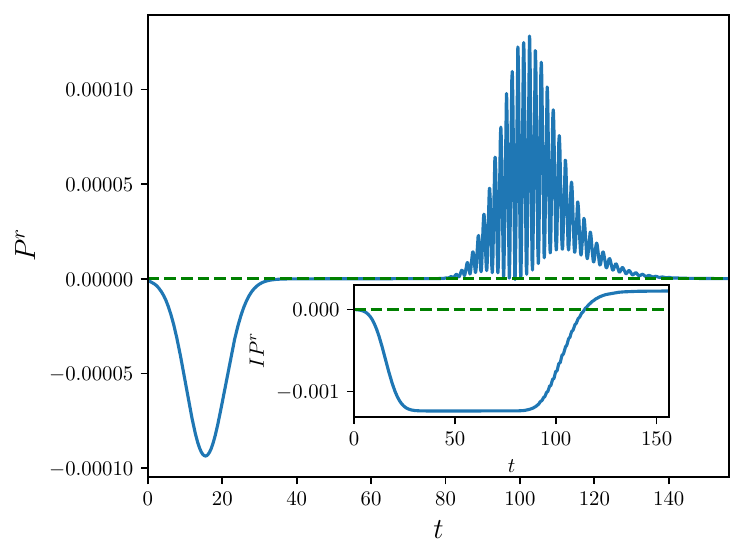}
      }
    \caption{Flux of energy $P^r$ and integrated energy flux $IP^r$ (insets) at the extraction point to $r_e = 30$ for (a) mini-boson star of Table~\ref{tab:spherical_cases} and (b) self-interaction boson star with $\omega_B=0.6\mu$. The parameters of the wavepacket are $r_0=40$ and $\omega=-2\mu$ for both cases and $\delta=10^{-2}$ for panel (a) and $\delta=10^{-3}$ for panel (b).}
    \label{fig:Jr}
\end{figure}

We allow the configurations listed in Table~\ref{tab:spherical_cases} to evolve, extracting the energy density $\rho$ and the radial flux $P^r$ at the point $r = r_e$. With this data, we can derive the amplification factors $\mathcal{A}_E$ and $\mathcal{A}_{tr}$. To illustrate the outcomes, we consider two representative cases: one corresponding to the free field scenario with $\delta = 10^{-2}$ and $\omega = 2\mu$ (i.e., $\omega_0 = 2.9\mu$), and another for the self-interacting case where $\omega_B = 0.6\mu$ with parameters $\delta = 10^{-3}$ and $\omega = -2\mu$. In Fig.~\ref{fig:Jr} we show $P^r(r=r_e)$ as a function of time. We also display the energy flux through the closed surface with radius $r_e$ over the time interval $[0,t]$ for those cases,
\begin{equation}\label{eq:IPr}
    IP^r(t) = 4\pi r_e^2\int_0^t dt' P^r(r=r_e,t') \,.
\end{equation}

\begin{figure}
    \centering
       \includegraphics[width=0.48\textwidth]{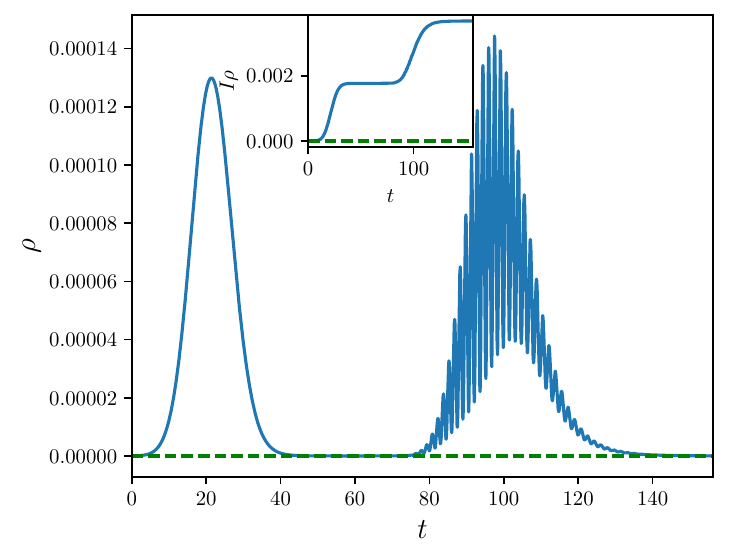}
    \caption{Energy density $\rho$ and integrated energy density $I\rho$ (inset) at the extraction point to $r_e = 30$ for the self-interaction boson star with $\omega_B=0.6\mu$ with the same values of the parameters as in Fig.~\ref{fig:Jr} panel (b).}
    \label{fig:rho}
\end{figure}

Then, the amplification factor can be obtained considering the minimum of $IP^r$ and the value at late times,
\begin{equation}
    \mathcal{A}_{tr} = \frac{IP^r(t\rightarrow \infty)-\min{IP^r}}{-\min{IP^r}} \, .
\end{equation}
In short, if we have that the final value of $IP^r$ is greater (less) than zero, then there was extraction (absorption) of energy and therefore amplification (attenuation). In a similar way we can extract the energy density amplification factor $\mathcal{A}_E$, computing the integral
\begin{equation}
    I\rho(t) = 4\pi r_e^2\int_0^t dt' \rho(r=r_e,t') \,.
\end{equation}
but now, the integral is monotonically increasing so we perform
\begin{equation}\label{eq:AE_extracted}
    \mathcal{A}_{E} = \frac{I\rho(t\rightarrow \infty)-{I\rho}_{\rm in}}{{I\rho}_{\rm in}} \, .
\end{equation}
where the quantity ${I\rho}_{\rm in}$ makes reference to the quantity $I\rho$ accumulated just after the wavepacket enters the star (in the intermediate plateau of the curve $I\rho(t)$). In Fig.~\ref{fig:rho} we display these quantities in the case of a self-interacting boson star. We will see in the following that both the density and the flux related observables $IP^r(t)$ and $I\rho(t)$ can lead to the correct determination of the amplification factors.

\begin{figure}
    \centering
    \subfigure[~~$\omega_B = 0.6\mu$]{
       \includegraphics[width=0.48\textwidth]{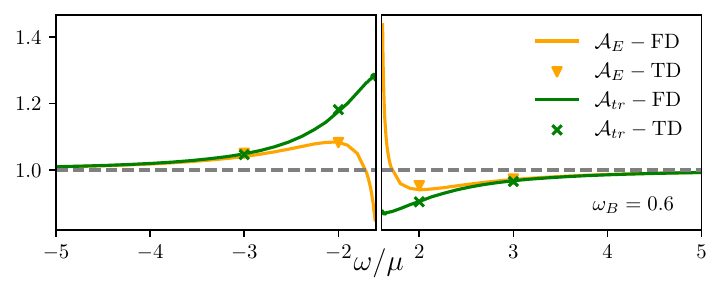}
      }
    \subfigure[~~$\omega_B=0.5\mu$]{
       \includegraphics[width=0.48\textwidth]{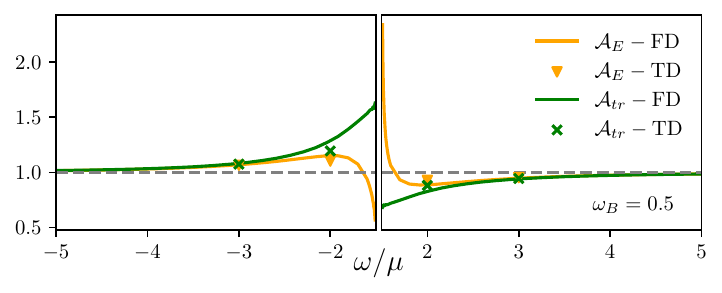}
      }
    \caption{Comparison between linear analysis and time-domain analysis for a wavepacket with $m=0$ scattering off spherical boson star. Here FD means frequency domain and TD means time domain analysis. An error bar was obtained according to a convergence test performed for the eight time domain points included in this figure, however the size of it is smaller than the width of the line of the frequency domain analysis.}
    \label{fig:factors_scan}
\end{figure}

Following the procedure described we extract $\mathcal{A}_E$ and $\mathcal{A}_{tr}$ from the set of around 60 evolutions of Table~\ref{tab:spherical_cases}. We arrive at the following conclusions:

\begin{itemize}
    \item Free field: Independently of the value of $\omega$, we obtain that whenever $\delta\ll1$ the amplification factors are consistent with one as expected. Increasing the amplitude, around values of $\delta=0.2$ we see a small deviation from $\mathcal{A}_E = \mathcal{A}_{tr} = 1$ corresponding to an increase of $\sim3.6\%$. For the case $\omega=-3\mu$ we slowly increased $\delta$ and saw that the amplification factors decreases monotonically up to the case $\delta =1$, where practically all the wave was absorbed by the boson star. Bigger values of $\delta$ lead to gravitational collapse of the system.
    \item Self-interacting field: Once again, both amplification factors are consistent with the linear results when $\delta$ is small. The agreement between the linear frequency-domain results and the nonlinear simulations is better for the configuration $\omega_B=0.6\mu$ than for the case $\omega_B=0.5\mu$ (notice that the last case is more compact than the former). This can be confirmed in Fig.~\ref{fig:factors_scan}. 
    
    We have used the time-domain configurations marked in Fig.~\ref{fig:factors_scan} to perform a convergence test and show that the difference between the two results is not related to lack of resolution. In particular we have used the three different resolutions $\Delta r = 0.05$, 0.025 and 0.0125 and a Courant factor of $1/4$. An error bar was obtained according to this convergence test, however the size of it is smaller than the width of the line of the frequency domain analysis. The interpretation is that the localized wavepacket description and the non-negligible gravitational backreaction play a relevant role in the time-domain analysis. Regarding the validity of the linear regime we obtain that for both boson star backgrounds and for the four values of $\omega$, when the amplitude is of order $\delta\sim10^{-2}$ we observe already appreciable differences with respect to the $\delta\ll1$ case. When $\delta$ reaches the order $10^{-1}$ we obtain that the boson star is perturbed with sufficient strength to collapse gravitationally in almost all of the cases independently of the value of $\omega$.
\end{itemize}

\begin{figure}
    \centering
       \includegraphics[width=0.48\textwidth]{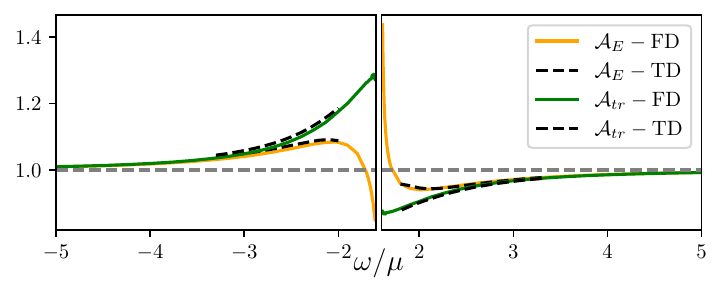}
    \caption{A ``continuum'' in $\omega$ for the factors $\mathcal{A}_E$ and $\mathcal{A}_{tr}$ using medium resolution $\Delta = 0.025$. Here FD means frequency domain and TD means time domain.}
    \label{fig:factors_continuum}
\end{figure}

Now that the parameter space of the system is recognized, we choose the case $\omega_B=0.6\mu$ and $\delta = 1\times10^{-3}$ to construct a continuum in $\omega$ for the factors $\mathcal{A}_E$ and $\mathcal{A}_{tr}$. This result is shown in Fig.~\ref{fig:factors_continuum}. 
We choose $\Delta r=0.025$ and also a Courant factor $\Delta t=0.25\times \Delta r$, place the wavepacket in each case initially at $r_{0}=40$, and we measure the data around $r_e=20$. We show the data of this scan in Fig.~\ref{fig:factors_scan}.  We observe that the data in perturbation theory and non-perturbation theory are consistent.

\subsection{Nonspherical cases}

We focus now in the cases where the star is nonspherical, the wavepacket is nonspherical, or both. 

\subsubsection{Nonspherical waves in a spherical background}

\begin{figure}
    \centering
       \includegraphics[width=0.49\textwidth]{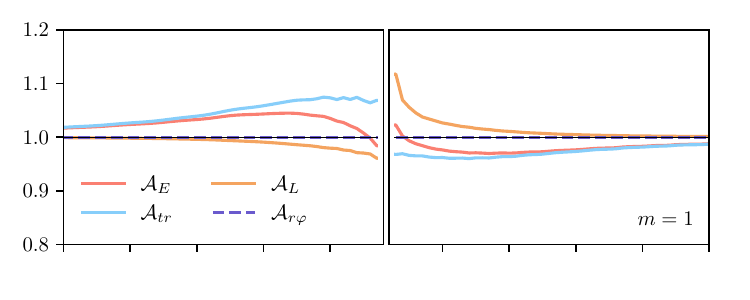}\\\vspace{-0.55cm}
       \includegraphics[width=0.49\textwidth]{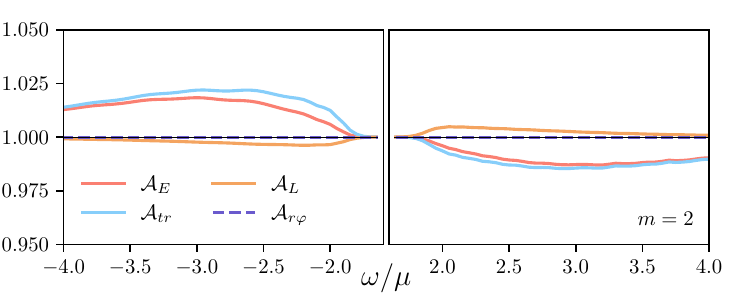}
    \caption{Amplification factors for a spherical background boson star, $m_B=0$, $\omega_B=0.6\mu$ but considering wavepackets with $m>0$. The amplification of angular momentum flux $\mathcal{A}_{r\varphi}$ coincides with $\mathcal{A}_N$.
    }
    \label{fig:factor_mB0_m1}
\end{figure}

We start by exploring some of the cases presented in the spherical configuration (Table~\ref{tab:spherical_cases}), now considering a nonzero azimuthal index for the wavepacket, $m_0 \neq 0$. In particular, we examine the self-interacting case with $\omega_B = 0.6 \mu$. For a spherically symmetric static star, angular momentum extraction is not possible. This is confirmed by Eq.~\eqref{eq:omega_pm}, which implies $m_+ = -m_-$, meaning that the definition of $\mathcal{A}_L$ in Eq.~\eqref{eq:Arphi} coincides with $\mathcal{A}_N$ in Eq.~\eqref{eq:AN}, so $\mathcal{A}_{r\varphi} = 1$ when $m_B = 0$. This does not apply to other amplification factors, especially for the angular momentum density factor. In Fig.~\ref{fig:factor_mB0_m1}, we corroborate this and present results for the static star with $\omega_B = 0.6 \mu$ for waves with $m = 1$ and $m = 2$ (the $m = 0$ case is shown in panel (a) of Fig.~\ref{fig:factors_scan}). We observe that energy amplification decreases as $m$ increases.
\begin{figure*}
    \subfigure[~~Wavepacket entering the star, $\Re(\Phi)$ in the equatorial plane.]{
    \hspace{-1.15cm}\includegraphics[height=0.14\textheight]{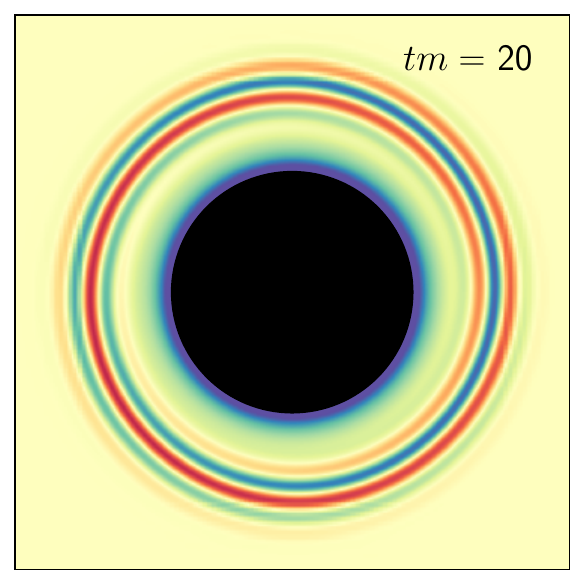}\hspace{-0.1cm}
    \includegraphics[height=0.14\textheight]{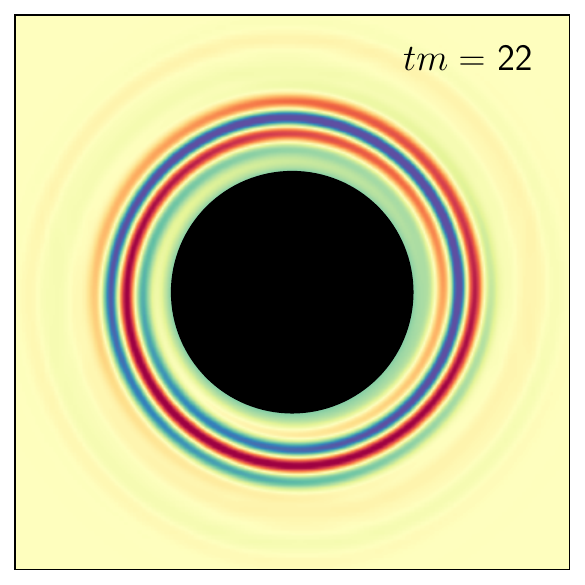}\hspace{-0.1cm}
    \includegraphics[height=0.14\textheight]{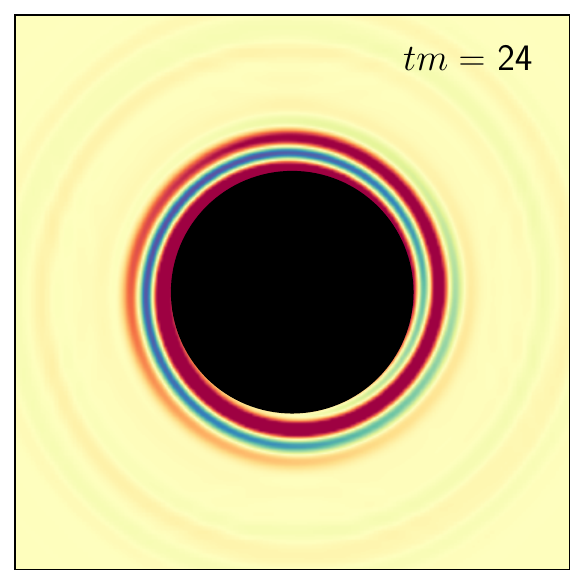}\hspace{-0.1cm}
    \includegraphics[height=0.14\textheight]{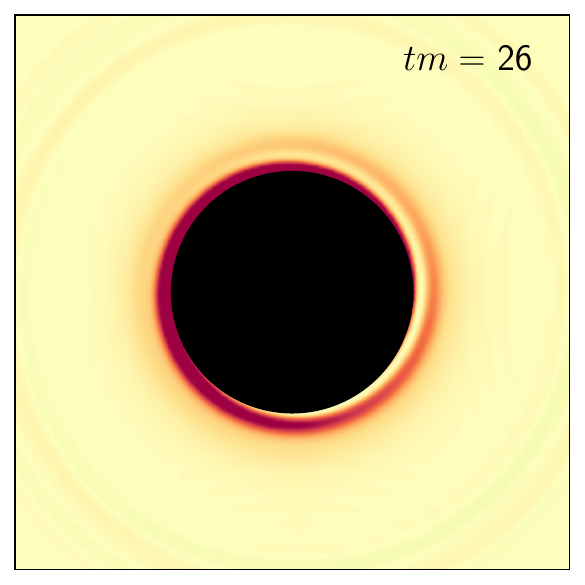}\hspace{-0.1cm}
    \includegraphics[height=0.14\textheight]{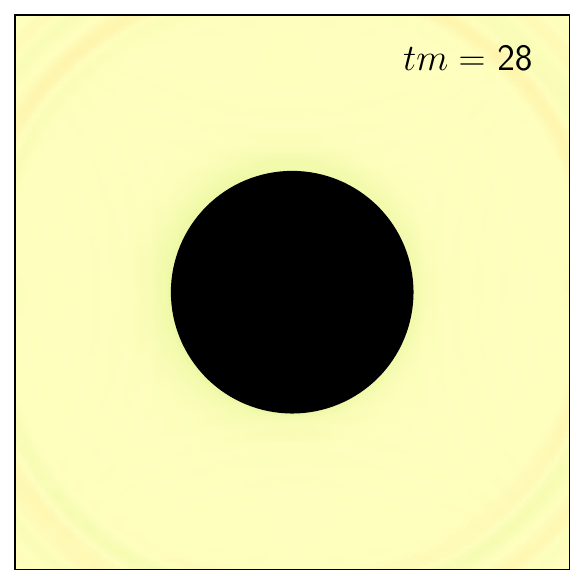}
    }
    \subfigure[~~Wavepacket leaving the star, $\Re(\Phi)$ in the equatorial plane.]{
    \includegraphics[height=0.14\textheight]{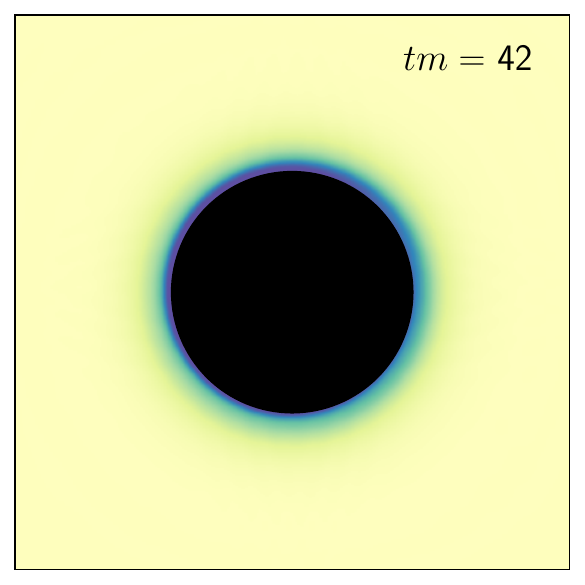}\hspace{-0.1cm}
    \includegraphics[height=0.14\textheight]{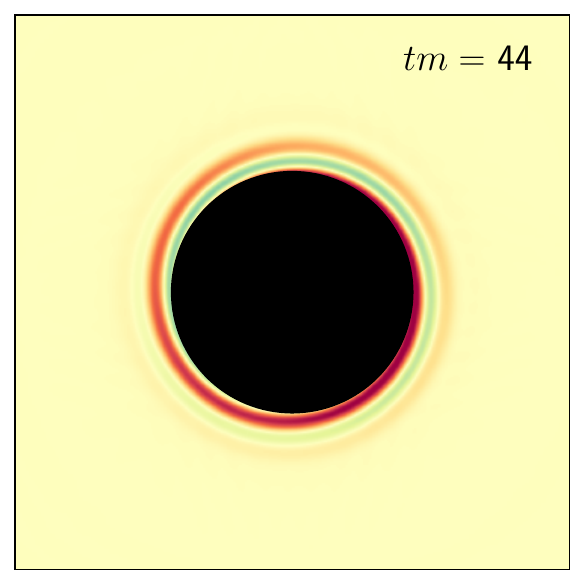}\hspace{-0.1cm}
    \includegraphics[height=0.14\textheight]{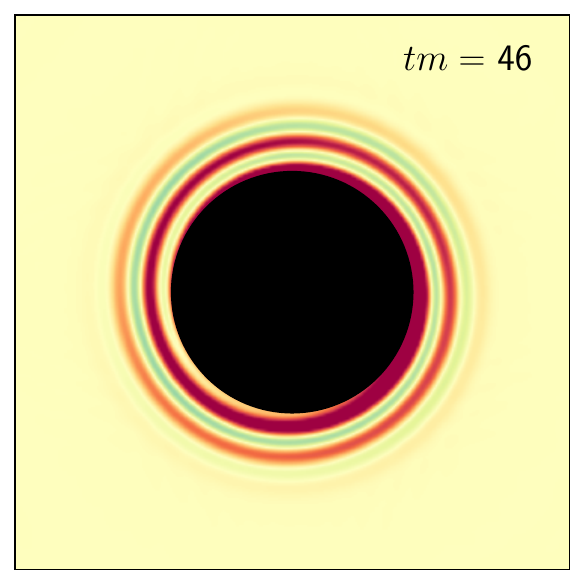}\hspace{-0.1cm}
    \includegraphics[height=0.14\textheight]{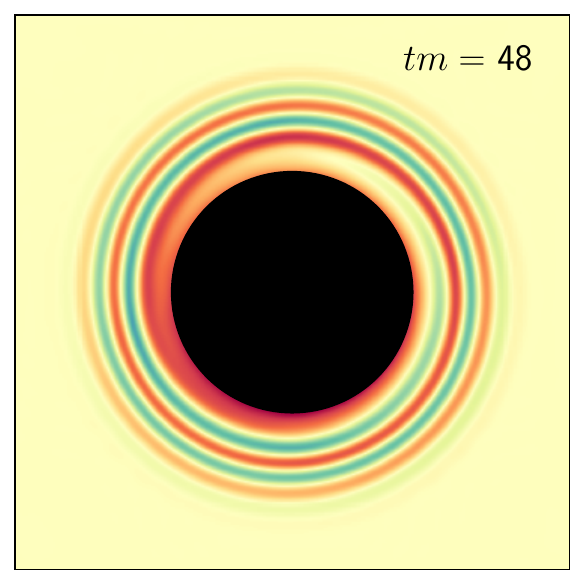}\hspace{-0.1cm}
    \includegraphics[height=0.14\textheight]{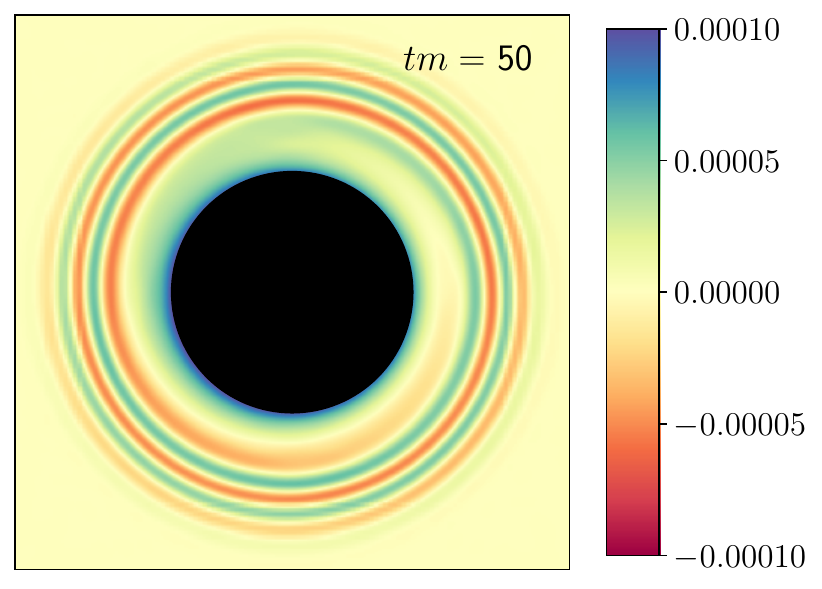}
    }
    \caption{Snapshots of the real part of the scalar field $\Phi$ at various times, illustrating the dynamics before and after a non-spherical wave with $m=1$ interacts with a spherical boson star. The central black region represents a radius of $2R_{99}$. The background star corresponds to the $\omega_B = 0.6\mu$ configuration listed in Table~\ref{tab:spherical_cases}, while the wavepacket properties are detailed in the main text. The box size is $30/\mu$.}
    \label{fig:snapshots_and_Pphi}
\end{figure*}

In contrast to the spherical case, the quantities $\rho$, $P^r$, $P^{\varphi}$, and $S_{\varphi r}$ now depend on $x$, $y$, and $z$. To extract the source terms, we select a sufficiently large radius and integrate over $d\Omega$ using the \texttt{Multipole} thorn. The amplification factors $\mathcal{A}_{E}$ and $\mathcal{A}_{tr}$ are then computed following the procedure in Eqs.~\eqref{eq:IPr}-\eqref{eq:AE_extracted}, with the substitutions:
\begin{equation}
\begin{split}
\rho(r=r_e)&\to \frac{1}{4\pi}\int d\Omega\rho|_{r_e}  \, , \\
P^r(r=r_e)&\to \frac{1}{4\pi}\int d\Omega P^r|_{r_e}\, .
\end{split}
\end{equation}
We apply a similar approach to calculate the angular momentum amplification factors $\mathcal{A}_{L}$ and $\mathcal{A}_{r\varphi}$. Specifically, the angular momentum flux factor is defined as
\begin{equation}
    \mathcal{A}_{r\varphi} = \frac{IS^{r\varphi}(t\to\infty) - \min\{s_{\omega_0} IS^{r\varphi}\}  }{-\min\{s_{\omega_0} I S^{r\varphi}\} }
\end{equation}
where 
\begin{equation}
    IS^{r\varphi}(t) = r_e^2  \int_0^t dt' \int d\Omega S^{r\varphi}(r=r_e,t') \, .
\end{equation}
Similarly, the amplification of angular momentum density is given by
\begin{equation}
    \mathcal{A}_L = \frac{I J^{\varphi}(t\to\infty) - IJ^{\varphi}_{\rm in}}{IJ^{\varphi}_{\rm in}}
\end{equation}
where 
\begin{equation}
I J^\varphi (t)= r_e^2  \int_0^t dt' \int d\Omega J^\varphi(r=r_e,t') \, ,
\end{equation}
and $IJ^{\varphi}_{\rm in}$ represents the total angular momentum density $IJ^{\varphi}$ just after the wavepacket crosses the extraction point $r_e$.

As a consistency check, we used a spherical background configuration and performed 3D simulations of a wavepacket with $m=1$, $\sigma_r=2.0/\mu$, $r_0=20/\mu$, and $\omega = 2.5\mu$, with a small amplitude of $\delta=1.65 \times 10^{-5}$. We then calculated the energy and angular momentum amplification factors. The results show good agreement with the linear approximations, with difference around $0.002$ for $\mathcal{A}_{E}$, $\mathcal{A}_{L}$, and $\mathcal{A}_{tr}$. For $\mathcal{A}_{r\varphi}$, the difference is approximately $0.015$, as this factor is more affected by noise. To illustrate the evolution of wavepackets with non-zero $m_0$, we include snapshots of the real part of $\Phi$ in the equatorial plane in Fig.~\ref{fig:snapshots_and_Pphi}.

\subsubsection{Nonspherical waves in a nonspherical background}

The choice of fiducial spinning boson star spacetimes results in distinct qualitative properties. This was observed in the linear case, as shown in Figs.~\ref{fig:factors_0p4} and \ref{fig:factors_0p45}, where amplification factors for the fiducial spinning boson stars are presented for $m=0$, 1, and 2. For example, energy amplification factors can reach values as high as 1.3, and the angular momentum density amplification factor can even become negative in certain regions of the parameter space.

The first four columns of Table~\ref{tab:spinning_cases} summarize the physical quantities of the background spinning configurations used in the nonlinear wavepacket scattering analysis. We add wavepackets with values $m = 0$, 1, and 2 (corresponding to $m_0$ in Eq.~\eqref{eq:Theta_3D} as 1, 2, and 3) and explore various values of $\delta$ and $\omega$. The selected wave parameters are shown in the last four columns of Table~\ref{tab:spinning_cases}.

In these simulations, we work in the perturbative regime, adding a wavepacket $\Theta$ from Eq.~(\ref{eq:Theta_3D}) to the boson star solution without solving the constraint equations or updating the background metric. We use a grid spacing of $\Delta x_i = 1.00/\mu$ within a box bounded by $x^i_{\max} = 60/\mu$, $x_{\min} = y_{\min} = -60/\mu$, and $z_{\min} = 0$, with 4 refinement levels and boundaries placed just beyond $30/\mu$. Reflection symmetry with respect to the $z=0$ plane is applied. For time evolution, we employ a Courant factor of 0.25. Additionally, for convergence studies, we test a grid spacing of $\Delta x_i = 0.80\mu$ with a fixed number of refinement levels, giving the innermost level a grid spacing of $\Delta x = 0.1\mu$. 

With the chosen settings, we were able to extract precise information about the transfer of energy and angular momentum between the wave and the boson star for the first fiducial model, characterized by $\omega_B = 0.45\mu$ (see Appendix~\ref{app:tests2} for further details). This configuration provided results with sufficient accuracy to analyze the amplification factors and validate the numerical approach. However, for the second fiducial model with $\omega_B = 0.4\mu$, we did not obtain results within the convergence regime, likely due to the increased compactness of the configuration or sensitivity of the system under these parameters. As a result, the subsequent discussion focuses exclusively on the first background configuration, where reliable and convergent results were achieved.

\begin{table}[b]
    \centering
    \begin{tabular}{cccc|c|cccc}
    \hline
     $\omega_B/\mu$ & $M\mu$ & $L\mu^2$ & $\phi_{\rm max}$ & $m$ & $\delta$                          & $\omega/\mu$ & $r_0\mu$ & $\sigma_r\mu$ \\ \hline
     0.45           &  0.666 & 1.056     & 0.0368           & $0,1$ & $[10^{-2},10^{-1}]$ & $\pm2,\pm2.5$ & 30 & 3 \\
     0.4            &  0.887 & 1.579     & 0.0367           & $0$ & $[10^{-2},10^{-1}]$ & $\pm2,\pm2.5$ & 30 & 3\\
    \hline
    \end{tabular}
    \caption{Configurations of spinning boson star with $m_B=1$ and $\sigma=0.05$ (solitonic) + wavepacket \eqref{eq:Theta_3D}. Instead of $\omega_0$, we report the quantity $\omega = \omega_0 - \omega_B$ for convenience in the analysis.}
    \label{tab:spinning_cases}
\end{table}

\begin{figure}
    \centering
    \subfigure[~~flux of energy]{
       \includegraphics[width=0.48\textwidth]{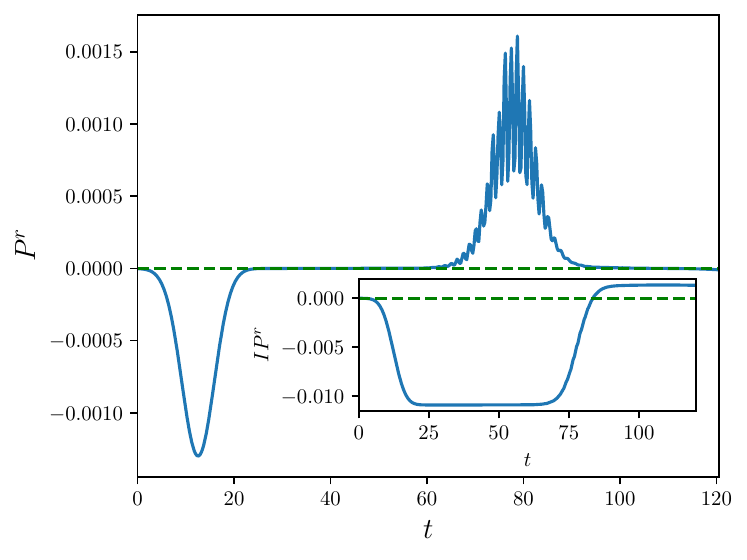}
      }
    \subfigure[~~flux of angular momentum]{
       \includegraphics[width=0.48\textwidth]{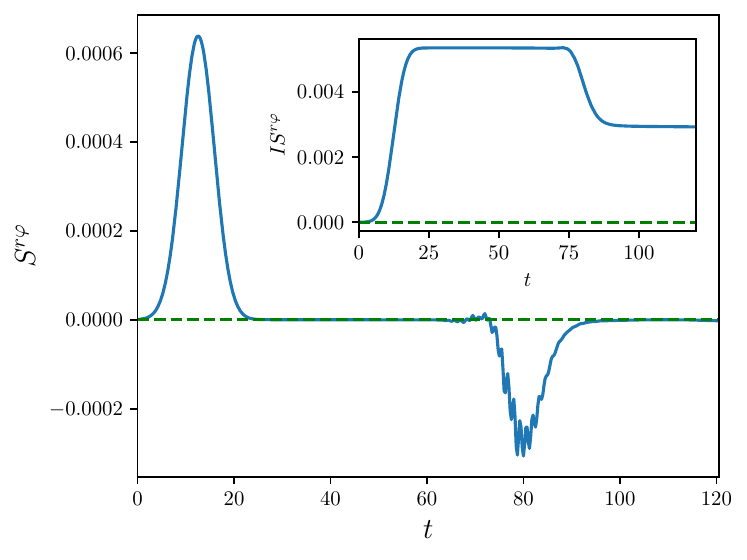}
      }
    \caption{Flux quantities in the scenario of a $m_0=1$ wavepacket with a spinning star in the background. The boson star correspond to the spinning boson star with $\omega/\mu=0.45$ in Table~\ref{tab:spinning_cases}. Panel (a): Flux of energy $P^r$ and integrated energy flux $IP^r$ (inset) at the extraction point to $r_e = 15/\mu$. Panel (b): Flux of angular momentum $S^{r\varphi}$ and the integrated angular momentum flux $IS^{r\varphi}$. The parameters of the wavepacket are $r_0=30/\mu$, $\omega=-2.5\mu$ and $\delta=0.1$.}
    \label{fig:Jr_spinning}
\end{figure}

To illustrate the system’s dynamics, Fig.~\ref{fig:Jr_spinning} shows the energy and angular momentum fluxes, along with their integrated values, for a representative case involving a wave and a $m_B = 1$ boson star with $\omega_B = 0.45\mu$. The wave has a frequency of $\omega = 2.5\mu$, and as expected, part of its energy and angular momentum is absorbed by the boson star. Consistently, the total angular momentum $L$, defined by the Komar integral \eqref{eq:globalquantJ}, stabilizes to a constant value shortly after the wavepacket exits the domain. This final value of $L$ is greater than that of the isolated boson star (see Table~\ref{tab:spinning_cases}), confirming that some angular momentum from the wavepacket was indeed absorbed by the star.\footnote{A similar result cannot be confirmed using the Komar mass integral Eq.~\eqref{eq:globalquantM}, as it does not settle by the simulation’s end, and extending the simulation is computationally prohibitive.} For the evolutions presented in Table~\ref{tab:spinning_cases}, we used wavepackets with parameters $\sigma_r \mu = 3$ and $r_0 \mu = 30$. Compared with Fig.~\ref{fig:spinning_contours}, we observe that the wave initially lies at twice the radius of the star in the equatorial plane, where the torus is wider.
\begin{figure}
    \centering

    \includegraphics[width=0.48\textwidth]{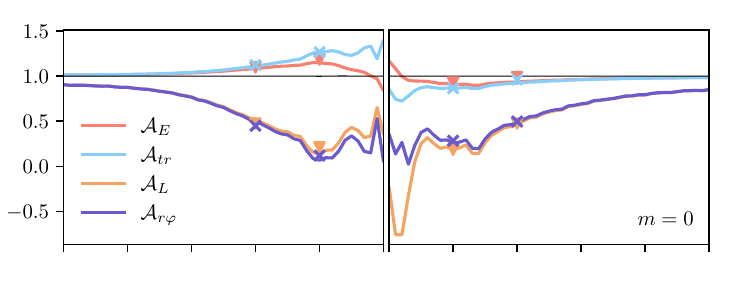}
    \includegraphics[width=0.48\textwidth]{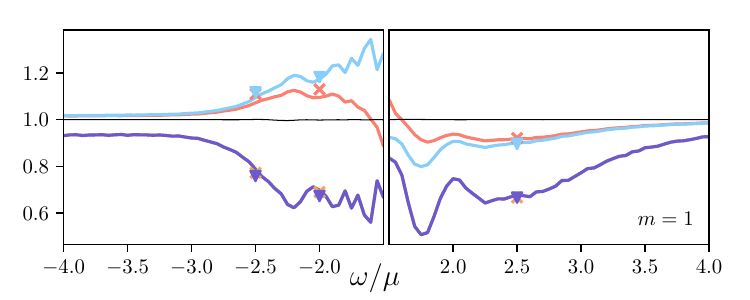}

    \caption{Comparison between linear analysis and time-domain analysis for a wavepackets with $m=0,m=1$ scattering off a spinning boson star with $\omega/\mu=0.45$ in Table~\ref{tab:spinning_cases}. The difference for the $m=0$ amplification factors have a maximum of the order of $0.065$, which is obtained for the $\omega=-2\mu$ wave. While for $m=1$ the maximum difference is 0.03.}
    \label{fig:factors_scan_rotating}
\end{figure}

Finally, in Fig.~\ref{fig:factors_scan_rotating}, we present the resulting amplification factors for waves with four different values of $\omega$, evolving in the background of the same spinning boson star, and compare these with the linear amplification factor obtained in Sec.~\ref{sec:linear}. The results show good agreement, with typical relative differences of the order of 1\%.

\subsection{Waves scattering inside a cavity}

In this section, we turn our attention to the problem of confining radiation within a cavity containing a boson star, with the goal of investigating potential instabilities that may arise. A similar study for the case of a flat-space Q-ball was performed in \cite{Cardoso:2023dtm}. The key question we aim to explore is whether the superradiance mechanism, can give rise to instabilities when the radiation interacts repeatedly with the star in the cavity. In cases of superradiance instability, the situation is straightforward: when radiation is confined, it undergoes multiple interactions with the amplifying body, which can lead to exponential energy growth and instability. However, for the superradiant mechanism discussed here, this behavior is not as readily anticipated.

The reason for this distinction lies in the nature of the scattered radiation. While the incoming radiation may be chosen to guarantee energy amplification, say by selecting the monochromatic wave \eqref{eq:Theta} and some negative value for the parameter $\omega<-\omega_B-\mu$, the scattered radiation emerges with the two frequencies $\omega_B+\omega$ and $\omega_B-\omega$, of which the second frequency component is not subject to the same energy-enhancing conditions. This introduces uncertainty regarding whether the system will exhibit instability in the same manner as other superradiant processes. Nevertheless, by confining radiation in a cavity with a boson star at the center, we can carry out numerical simulations to explore its evolution and determine whether any instability manifests under such conditions.

To simulate the cavity, we place a spherical mirror at $r = r_m$, where we impose Dirichlet boundary conditions:
\begin{equation}
    \Phi(r_m) = 0 \, ,
\end{equation}
as the reflective boundary. Previous studies \cite{Sanchis-Gual:2016tcm, Sanchis-Gual:2015lje, Dailey:2023mvn} have already applied numerical relativity techniques to evolve waves within a cavity in the presence of gravity. Specifically, in \cite{Sanchis-Gual:2016tcm, Sanchis-Gual:2015lje}, this was explored in the context of superradiance in the Reissner-Nordstr\"om black hole, where the presence of a charged scalar field led to a superradiant instability (black hole bomb). They found that this process could extract energy from the black hole, but the instability \textit{ceased} before the black hole’s charge was entirely depleted.

\begin{figure}
    \centering
    \subfigure[~~integrated energy]{
       \includegraphics[width=0.47\textwidth]{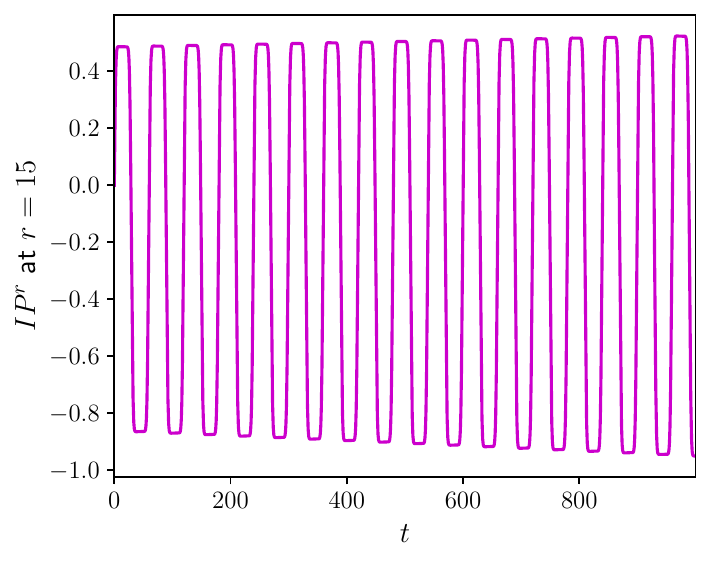}
      }
      \subfigure[~~change of the conserved charge]{
       \includegraphics[width=0.47\textwidth]{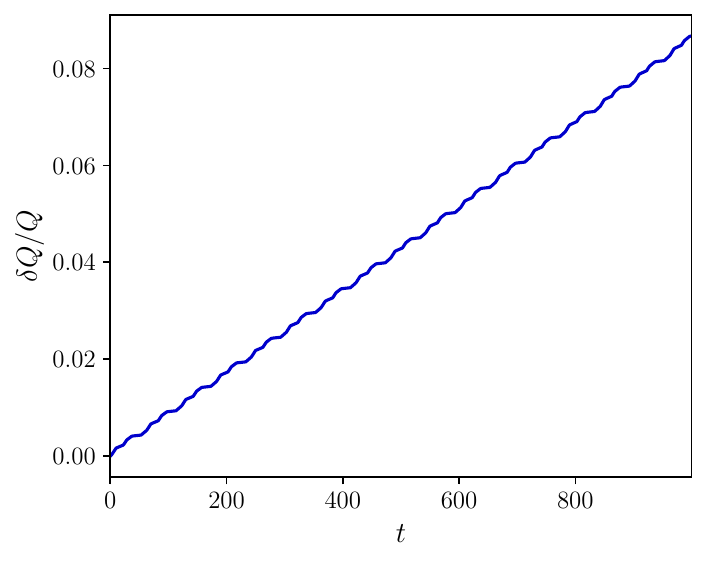}
      }
    \caption{Test of the scalar sector in a cavity. Massless scalar field ($\mu=0$) in Minkowski spacetime reflecting at a mirror placed at $r_m=30$. In panel (a) the time-integrated flux $IP^r$ was extracted at $r_e=15$. Panel (b) shows consistency of the implementation. }
    \label{fig:vac_test}
\end{figure}

In the following we will implement in a natural and simplified way the scenario of a spherically symmetric wavepackets interacting with a spherically symmetric boson star in a cavity. For this, we will use the evolution code described in Sec.~\ref{sec:evol_scheme_sph}. To verify the reliability of the implementation, we tested the dynamics of a simple scalar wave ($V(\Phi) = 0$) in a fixed Minkowski spacetime, as well as free wavepackets within the theory described by Eq.~\eqref{eq:action}. The results are plotted as Fig.~\ref{fig:vac_test}, where we consider the flux of energy $P^r$ through the fixed spherical surface with radius $r_e=15$ in a cavity with mirror placed at $r_m=30$. 
After validating the code, we proceed to study the scenario of radiation trapped in a cavity containing a boson star.

We begin with nonlinear evolutions of a test scalar field, initially prescribed by Eq.~\eqref{eq:fullPhi}, in the fixed (non-dynamical) geometry of a boson star solution. Next, we consider the fully self-consistent case, allowing the geometry to evolve in response to the combined dynamics of the boson star and wavepacket with very small value of the amplitude $\delta$. We focus for that on the (spherically symmetric) self-interacting model, with background solutions characterized by the frequency $\omega_B = 0.6\mu$, corresponding to the first row in Table~\ref{tab:spherical_cases}. In this section we have changed to Schwarzschild coordinates. In this coordinates the initial data is different to the one presented in Sec.~\ref{sec:nonlinear_id}. We start from~\eqref{eq:spherical_momentumConstraint} and~\eqref{eq:spherical_Hamiltonian} choosing $K_b=0$ and $b=1$. So instead of Eqs.~\eqref{eq:hamiltonian} \eqref{eq:momentum}, we obtain
\begin{align}
\label{eq:spherical_momentumConstraint2}
        &{{\mathcal{M}_r=-\frac{2}{r}K+{8\pi P_r}=0}} \, ,\\ 
\label{eq:spherical_Hamiltonian2}
    \begin{split}
        &\mathcal{H}=\frac{1-a}{r}-D_a+8\pi r a \rho=0 \, .
    \end{split}
\end{align}
After solving those equations, we obtain the initial data for the simulations.

Since we are simulating a cavity confined to a limited spatial region, the influence of spacetime curvature on energy flux measurements must be considered, especially if we anticipate wave energy amplification. To assess this, we employ two closely related quantities to estimate the energy extracted from the star. The first quantity is a generalization to curved spacetime of the total energy flux through the spherical surface $\mathcal{S}$ of radius $r_e$ between $t = 0$ and $t$, denoted by $IP^r$ in Eq.~\eqref{eq:IPr}:
\begin{equation}\label{eq:curved_IPr}
IP^r = \int_0^\tau \oint_\mathcal{S} P^i s_i \, d\mathcal{S} \, d\tau = 4\pi r_e^2 \int_0^t P_r \left(\frac{\alpha b}{\sqrt{a}}\right) dt',
\end{equation}
where $s^i$ is the spatial normal to $\mathcal{S}$ and $\tau$ is the proper time as measured by the static observer. This quantity applies in a dynamic spacetime, given in Eq.~\eqref{eq:metric_spherical}. Alternatively, we can use the conserved current in static spacetimes, associated with the Killing vector $K = \partial_t$, to estimate the energy variation within $\mathcal{S}$. Integrating the conserved current $J_E^\mu := T^{\mu \nu} K_\nu$ over a 3-surface at constant $r = r_e$  gives
\begin{equation}\label{eq:curved_CPr}
CP^r = \int J_E^\mu s_\mu \, d\epsilon = 4\pi r_e^2 \int_0^t J_E^r \left(\alpha \sqrt{a} b\right) dt' \, ,
\end{equation}
where $d\epsilon$ is the 3-surface element, which for the line element in Eq.~\eqref{eq:metric_spherical} is  $d\epsilon = \alpha b r^2 \sin \theta \, d\theta \, d\varphi \, dt'$. In static spacetimes, $CP^r$ vanishes (the closed integral of $J_E^\mu$ over a the 3-dimensional boundary of a 4-dimensional volume is equal to zero and the integral over a $t$ constant 3-surface yields the mass). Applying $CP^r$ to dynamic cases thus indicates energy variation within the system.

The definitions of $IP^r$ and $CP^r$ differ slightly. Here, $P_r$ denotes the radial (covariant) component of energy flux measured by an observer with 4-velocity $n^\mu$, while $J_E^r$ is just the radial component of the 4-vector $J_E^\mu$. Direct calculation shows they are related by  $J_E^r = \alpha P_r / a$, leading to  $CP^r = 4\pi r_e^2 \int_0^t P_r \left(\alpha^2 b / \sqrt{a}\right) dt$, which has an additional factor of $\alpha$ in the integrand compared to Eq.~\eqref{eq:curved_IPr}.

\begin{figure}
    \centering   
    \includegraphics[width=0.48\textwidth]{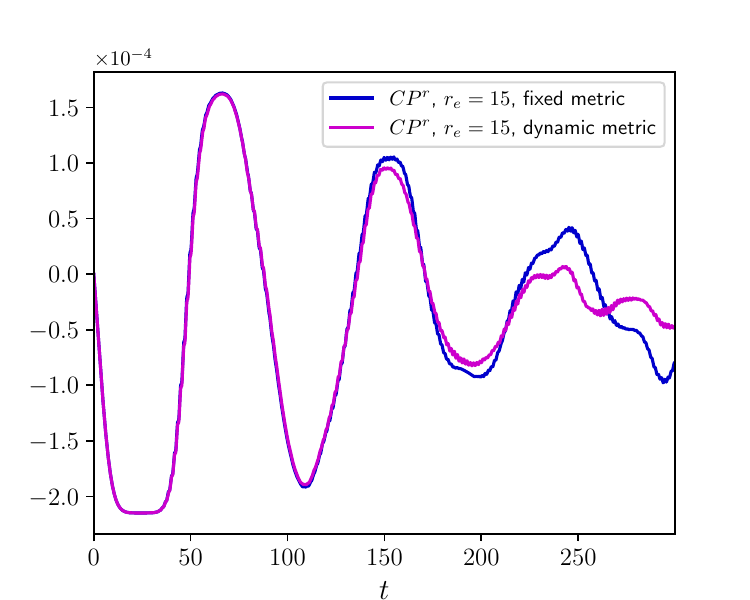}
    \includegraphics[width=0.48\textwidth]{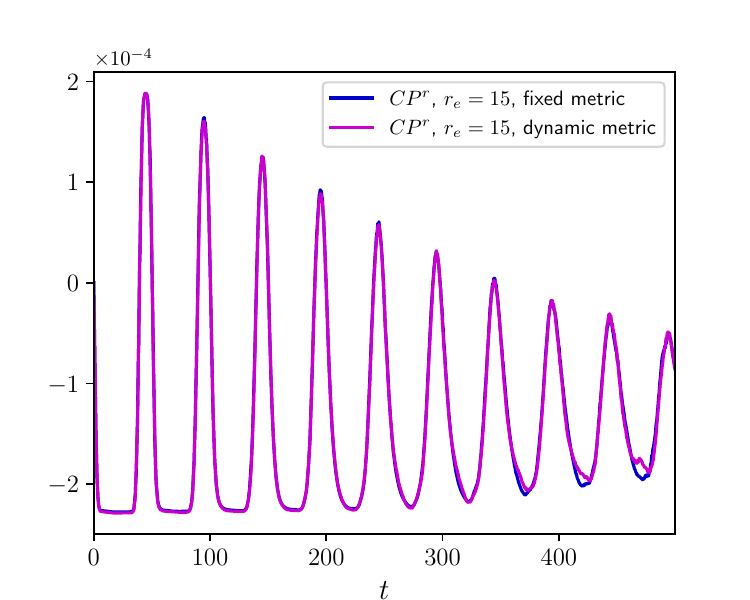}
    \caption{Integrated energy for the case $\delta = 10^{-3}$. The upper panel corresponds to $\omega = -2\mu$, while the lower panel shows results for $\omega = -3\mu$. Both definitions of total energy flux, $IP^r$ and $CP^r$ [Eqs.~\eqref{eq:curved_IPr} and \eqref{eq:curved_CPr}], are found to coincide. Across all four evolutions presented, the total Noether charge remains within 1\% of its initial value.}
    \label{fig:cav_delta-3}
\end{figure}

\begin{table}[b]
    \centering
    \begin{tabular}{ccccc}
    \hline
        $\delta$ & $\omega/\mu$ & $r_0\mu$ & $\sigma_r\mu$ & $r_m \mu$\\ \hline
        $10^{-3}$ & $-2$ & 15 & 5 & 30\\
        $10^{-3}$ & $-3$ & 15 & 5 & 20\\
        $5\times10^{-2}$ & $-3$ & 10 & $\sqrt{5}$ & 20\\
    \hline
    \end{tabular}
    \caption{Chosen parameters for the waves scattering inside a cavity. For all these cases the boson star correspond to the spherically symmetric self-interacting boson star in the first row of Table~\ref{tab:spherical_cases}.}
    \label{tab:cavity_cases}
\end{table}
We explored various cases of $\omega$ and $\delta$. To ensure stability and efficiency in long-term simulations, we adjusted some parameters, as listed in Table~\ref{tab:cavity_cases}. The cases with an amplitude of $\delta = 10^{-3}$ represent a perturbative scenario (see, for example, Fig.~\ref{fig:Jr}). We conducted simulations with both fixed and dynamic metrics to test the program and achieved consistent results in both setups for these perturbative cases: specifically, for $\omega = -2\mu$ with the mirror placed at $r_m = 30/\mu$ and $\omega = -3\mu$ with $r_m = 20/\mu$. Results are shown in Fig.~\ref{fig:cav_delta-3}.

Initially, we observe an expected amplification in energy, consistent with the linear analysis in Sec.~\ref{sec:linear}. However, after a few reflections, no additional energy is extracted. Both the fixed and dynamic metric cases yield the same qualitative conclusions.

\begin{figure}
    \centering
    \includegraphics[width=0.49\textwidth]{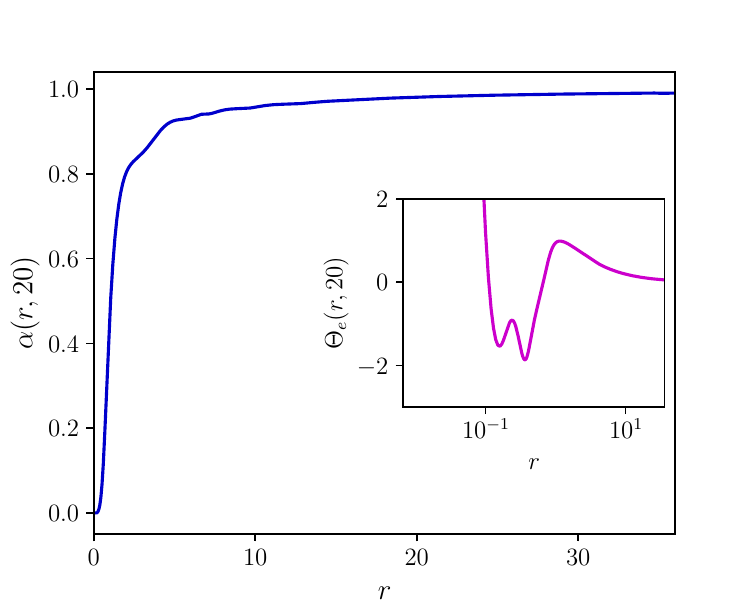}
    \caption{Lapse and expansion at $t=20$ in $\delta=5\times10^{-2}$ case.}
    \label{fig:blackhole}
\end{figure}

We also conducted simulations using a wave with an amplitude 50 times greater than in the perturbative cases, making it distinctly non-perturbative. In this higher-amplitude scenario, the dynamics are significantly altered: the strong ingoing wavepacket leads to gravitational collapse, ultimately resulting in the formation of a black hole, as shown in Fig.~\ref{fig:blackhole}. The first figure shows the lapse near the origin approaches to 0, which is a result of the singularity avoiding gauge. Time freezes near the singularity. The inset shows the expansion of the outgoing null geodesics $\Theta_e$, given by $\Theta_e=(2/r+D_b)/\sqrt{a}-2K_b$. If $\Theta_e$ vanishes for any given $r$, then there is an apparent horizon located there. This outcome highlights the nonlinear effects that arise with increased wave amplitude, in contrast to the milder, oscillatory behavior observed in the perturbative regime. According to these results there must be a critical amplitude threshold beyond which the system’s response becomes dominated by gravitational self-interaction, leading to an irreversible collapse. 

\section{Conclusions}\label{sec:conclusions}

In this work, we have studied the phenomenon of superradiance in the context of 
spherical and (axisymmetric) rotating boson stars with a self-interacting scalar. By combining linear perturbative analyses with nonlinear time-domain simulations, we have analyzed the wave amplification and its connection to the dynamics of these configurations.  

One of the key findings of this study is the good agreement between the results obtained from the perturbative and the nonlinear approaches. In particular, the amplification factors for energy and angular momentum show consistency across both methods when the amplitude of the incident wave is small. Remarkably, in some scenarios, the energy of the outgoing wave exceeds that of the ingoing wave by up to 40\%, making the superradiant processes in this system efficient. This results are made possible by including nonlinear interactions in the scalar field potential, which on the other hand is a known stabilization mechanism for rotating boson stars. In the simpler free field case, rotating boson stars are prone to non-axisymmetric instabilities. The inclusion of nonlinear terms in the potential (similar to the Q-ball scalar potential) not only stabilizes these objects but also enables the superradiance process to happen \cite{Saffin:2022tub}. This stabilization is a requirement for understanding the long-term behavior of such configurations and their interaction with perturbations.  

Our analysis of spherically symmetric configurations also demonstrates the precise conditions under which energy amplification occurs, highlighting the consistency between perturbative analyses in momentum space \cite{Gao:2023gof} and real-time simulations. By employing both the perturbative and nonlinear approaches, we confirm that these stars act as robust scatterers, enabling energy transfer even in the absence of real-space rotation. The investigation of wave scattering inside a cavity containing a boson star provided additional information about superradiance. The cavity setup allows for the confinement of scalar waves, and in particular allowed to conclude that no superradiant instabilities, which is more stringent than superradiant amplification, seem to develop, although our investigation in this regard is not exhaustive.

The study also demonstrates the capability of spinning self-interacting boson stars to transfer energy effectively to incident waves under specific conditions. The nonlinear potential of these stars, as in the spherical background scenario, amplifies even small wave perturbations, allowing for efficient energy extraction. From a methodological perspective, the implementation of a two-dimensional spectral solver and a three-dimensional numerical relativity framework proved to be robust and reliable. These tools successfully handled both spherical and non-spherical configurations, with convergence studies and analytical comparisons confirming their accuracy.  

Overall, this work provides some new insights into the energy extraction of generic boson stars, and reinforces the significance of these exotic objects in scalar dark matter models and their potential as a test bed for exploring strong-gravity phenomena and nonlinear field dynamics. Future research could expand upon these findings by exploring setups or modifications that can realize superradiant instabilities and examining the interaction of boson stars with electromagnetic radiation, where superradiance—if present—might result in a continuous mass loss for an illuminated boson star, say, in an astrophysical binary system. 

\acknowledgments
We thank Paul Saffin, Qi-Xin Xie and Guo-Dong Zhang for helpful discussions. We acknowledge support the National Key R\&~D Program of China under grant No.~2022YFC2204603 and from the National Natural Science Foundation of China under grant No.~12075233 and No.~12247103.

\appendix

\section{Convergence tests}\label{app:tests}

\begin{figure*}
    \centering
       \includegraphics[width=0.9\textwidth]{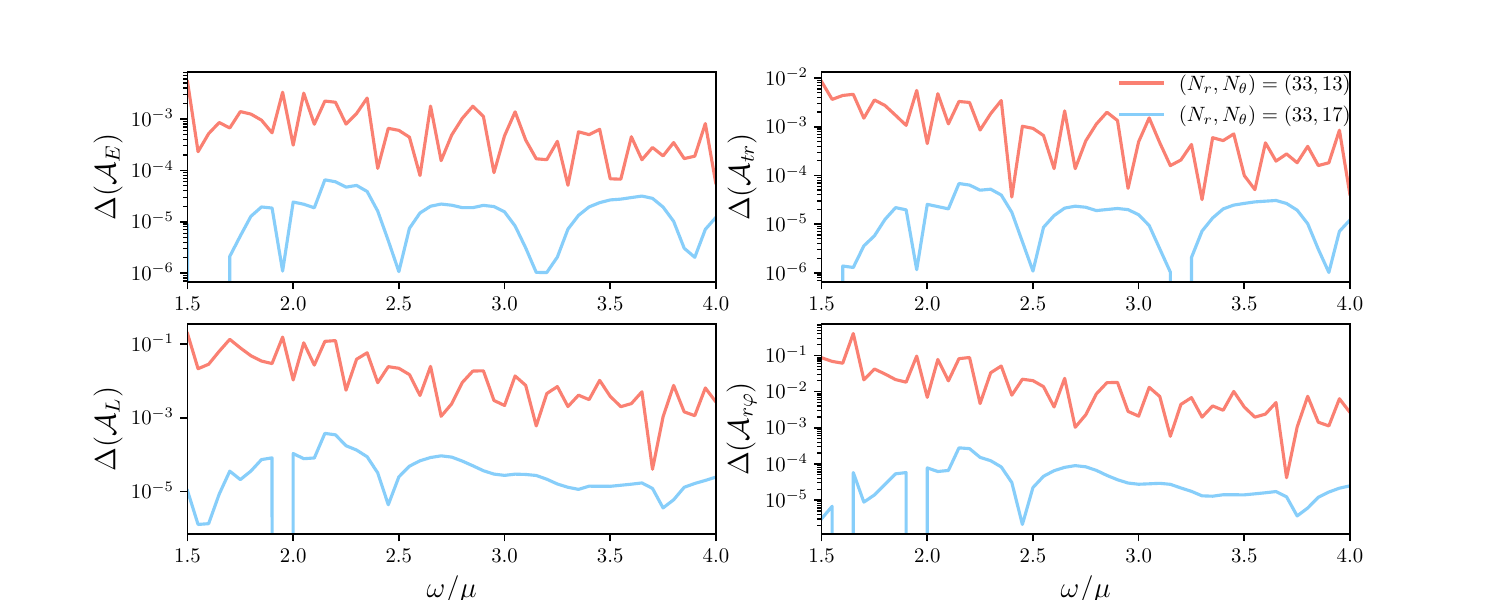}
    \caption{The relative differences in the amplification factors $\Delta(\mathcal{A})$ compared to the higher-resolution reference case using $(N_r,N_{\theta})=(33,21)$. Here we show results for the $\omega>\mu+\omega_B$ cases, however similar results hold in the $\omega<-\mu-\omega_B$ region.}
    \label{fig:Ntheta_linear}
\end{figure*}

In this appendix, we present resolution studies for both the calculation of the linear amplification factors and the corresponding time-domain evolutions in 3D. These tests are conducted using the background of a spinning boson star with $\omega_B = 0.45\mu$ and a solitonic potential characterized by the parameter $\sigma = 0.05$. The global properties of this particular solution were provided in Table \ref{tab:spinning_cases}.

\subsection{Frequency-domain}\label{app:tests1}
\begin{figure*}
    \centering
       \includegraphics[width=0.82\textwidth]{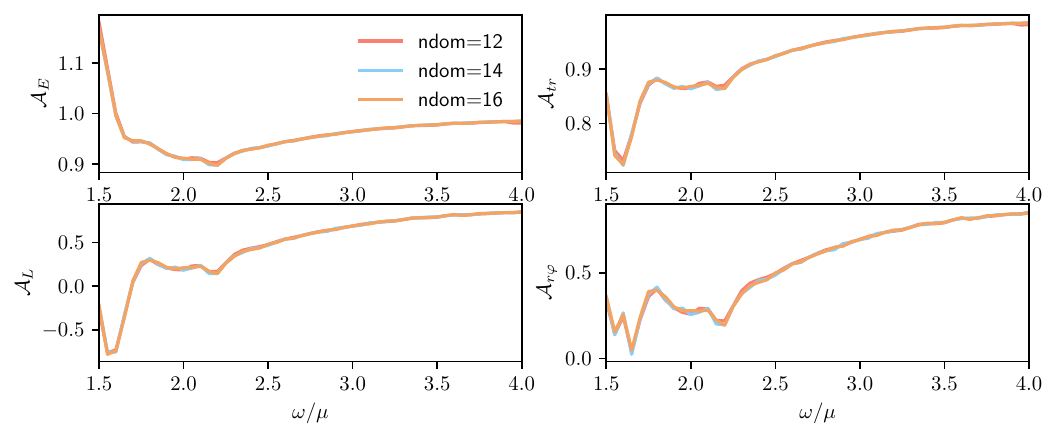}
    \caption{The comparison of amplification factor about increasing the number of domains, $n_{\rm dom}$. Here we show results for the $\omega>\mu+\omega_B$ cases, however similar results hold for the $\omega<-\mu-\omega_B$ region.}
    \label{fig:ndom_linear}
\end{figure*}

To validate the code used in Sec.\ref{sec:linear} for solving the linear coupled partial differential equations in Eqs.~\eqref{EOM_etapm}, we conduct a series of convergence tests. The implementation uses the \texttt{Kadath} library, which provides a robust framework for spectral methods. Specifically, we employ Chebyshev polynomials as the spectral basis and utilize the library’s \textit{polar space} module. This module operates in spherical coordinates under the assumption of axisymmetry, dividing the computational domain into spherical shells and compactifying the outermost shell to efficiently handle boundary conditions at infinity.

This approach is similar to the method used to construct spinning boson star solutions in Sec.~\ref{sec:stationary}, where asymptotic boundary conditions \eqref{eq:out_bc} are imposed exactly because $r \to \infty$ lies within the computational domain. However, a significant difference exists in the asymptotic behavior of the fields. While the boson star field $\phi$ decays exponentially to zero at infinity, the perturbative fields $\eta^{\pm}$ oscillate with a fixed wave number and decay as $1/r$. This oscillatory behavior makes compactification less suitable for the perturbative system. Instead, we solve the system on a domain extending to a large but finite outer boundary, where we impose the boundary condition \eqref{eq:eta_outBC}.

Since the equations are linear, the solver converges after a single iteration. To test the numerical accuracy, we vary the number of Chebyshev spectral coefficients in the radial dimension, $N_r$, and in the angular dimension, $N_\theta$. For the results presented, we used $N_r = 33$, $N_\theta = 17$, and a total of $n_{\rm dom} = 14$ radial domains with boundaries at $r\mu = 1$, 2, 4, and $8m$ for $1 \leq m \leq 10$.

For convergence testing, we first fixed $N_r$ and $n_{\rm dom}$ and then varied $N_\theta$, using $N_\theta = 13$ (lower resolution) and $N_\theta = 21$ (higher resolution). Fig.~\ref{fig:Ntheta_linear} shows the results of this test. The relative differences in the amplification factor compared to the higher-resolution reference case consistently approach zero, confirming convergence and the reliability of the implementation.

Similarly, we tested the impact of varying the radial spectral resolution by considering values of $N_r < 33$, while keeping the number of domains $n_{\rm dom}$ and angular coefficients $N_\theta$ fixed. We observed that the differences relative to the $N_r = 33$ case converge to zero far from the mass gap. However, near the mass gap, the relative differences saturate due to round-off errors. A more noticeable improvement in accuracy is achieved by increasing the number of domains, $n_{\rm dom}$. To illustrate this, we included tests with a lower resolution $n_{\rm dom} = 12$ and a higher-resolution reference case $n_{\rm dom} = 16$ (adding radial domains with boundaries at $r\mu=88$ and 96). The results, shown in Fig.~\ref{fig:ndom_linear}, demonstrate convergence to zero. Furthermore, the particle number amplification factor $A_N$ consistently converges to 1 across all convergence studies, confirming the robustness of the implementation.

\subsection{Time-domain}\label{app:tests2}

\begin{figure}
    \centering
       \includegraphics[width=0.45\textwidth]{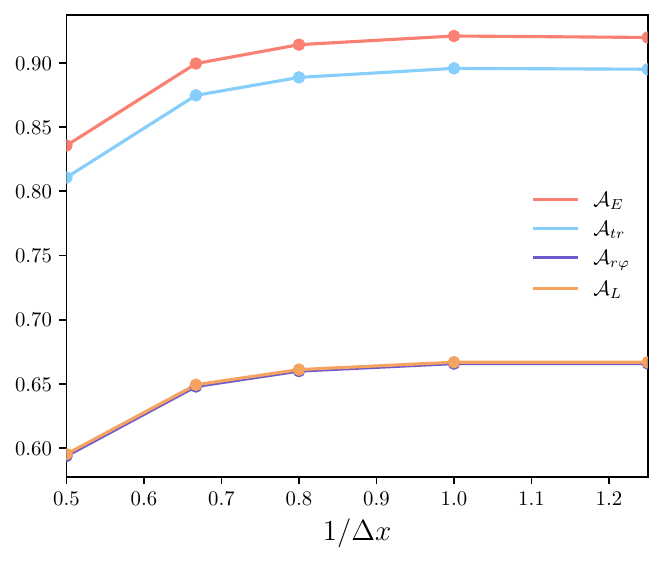}
    \caption{Convergence of amplification factors with spatial resolution $\Delta x_i$. For convenience, we choose the horizontal axis to be $1/\Delta x$.}
    \label{fig:time_conver}
\end{figure}

Throughout the paper, we have consistently ensured the reliability of our results by analyzing them with at least two resolutions to verify convergence. For a more detailed convergence study, we focused on the $m=1$ ($m_0=2$) wavepacket case, characterized by parameters $\sigma=3.0/\mu$, $\delta=0.1$, and centered at $r_0=15$, propagating in the same spinning boson star background as in the test presented in Sec.~\ref{app:tests1}.

To evaluate convergence systematically, we repeated the evolution while varying the spatial resolution $\Delta x_i = \{0.8, 1.0, 1.25, 1.5, 2.0\}$. The resulting values of the four amplification factors, plotted as a function of $1/\Delta x_i$, are presented in Fig.~\ref{fig:time_conver}, demonstrating the expected convergence behavior.

\bibliographystyle{unsrt} 
\bibliography{ref}

\end{document}